\documentclass[preprint,12pt]{elsarticle}

\usepackage{amssymb}
\usepackage{graphicx}
\usepackage{dcolumn}
\usepackage{bm}
\usepackage[utf8]{inputenc}
\usepackage[T1]{fontenc}
\usepackage{mathptmx}
\usepackage{chemformula}
\let\ce\ch
\usepackage{caption}
\usepackage{subcaption}
\usepackage{natbib}
\usepackage{float}
\usepackage{multirow}
\usepackage{array}
\usepackage{tabularx}
\def \mycolor {black}

\journal{Journal of Computational Physics}

\begin{document}

\begin{frontmatter}



\title{Load-Balanced Neural-Network-Aided Adaptive Tabulation of Phase Equilibria for Transcritical Multiphase Flows}

\author[1]{Navneeth Srinivasan}
\author[1]{Po-Yu Lai}
\author[1]{Yousef Eldahshoury}
\author[1]{Suo Yang \corref{cor1}}

\affiliation[1]{organization={Department of Mechanical Engineering, University of Minnesota -- Twin Cities},
    addressline={111 Church Street SE}, 
    city={Minneapolis},
    postcode={MN 55455}, 
    country={USA}}
\cortext[cor1]{Corresponding author (suo-yang@umn.edu)}

\begin{abstract}
Simulating high-pressure transcritical flows with phase change is critical for advanced propulsion, but accurately resolving these flows remains challenging in thermodynamically complex regimes. These conditions are commonly treated using vapor-liquid equilibrium (VLE) modeling, which can dominate runtime and suffer from convergence failure. Standalone machine learning models (Artificial Neural Networks, ANN) incur high evaluation cost, while adaptive tabulation (In-Situ Adaptive Tabulation, ISAT) retains convergence risks associated with fragile iterative solvers. This study introduces an integrated ISAT-ANN framework designed to overcome the complementary limitations of standalone neural networks and adaptive tabulation, with the ANN eliminating fragile iterative solves during table expansion and ISAT avoiding the cost of repeated network evaluation. To translate these improvements to large-scale simulations, an $O(N)$ greedy dynamic load-balancing algorithm redistributes nonuniform thermodynamic workloads across processors, while adaptive mesh refinement concentrates resolution around the droplet interface and regions of deformation. Together, these components form an integrated framework addressing the coupled challenges of thermodynamic robustness, parallel scalability, and spatial computational cost, enabling high-fidelity resolution of strongly nonuniform transcritical flows. Tested on a nominal Mach 5 shock-droplet interaction benchmark across 16 central processing unit (CPU) cores and an adaptive mesh refinement (AMR) mesh peaking at 190,482 cells, the framework successfully completed the reported simulations and resolved complex shock-interface dynamics without solver failures. It achieved a 2.30$\times$ speedup in thermodynamic calculations on a single core and 2.62$\times$ in parallel over standalone ANNs, while AMR provided a further 2.18$\times$ speedup in total and a 2.92$\times$ speedup in VLE computation relative to the static-mesh.

\end{abstract}





\end{frontmatter}


\section{Introduction}
To meet the increasing demand for high-performance combustors, operating chamber pressures are often pushed into the supercritical regime. Under these elevated pressures, injected multi-component liquid propellants and fuel-air mixtures undergo complex transcritical phase changes during spray breakup, vaporization, mixing, and combustion \cite{jofre2021transcritical,fathi2022large}. Efficient spray breakup and vaporization are critical to minimizing the effective vaporization time and optimizing combustor performance \cite{faeth1977current}. However, conventional liquid injectors and multi-component fuel formulations optimized for low-pressure regimes perform suboptimally at high pressures, necessitating comprehensive redesign and optimization \cite{chehroudi2012recent}.

Understanding transcritical and supercritical mixing and combustion processes requires high-fidelity computational fluid dynamics (CFD) modeling capable of accounting for non-ideal real-fluid behavior. Near the critical point, fluids exhibit extreme variations in thermophysical properties, such as density and specific heat, making accurate property prediction essential yet computationally challenging \cite{oschwald2006injection}. Because minor variations in temperature and pressure severely alter fluid structure near the critical point, modeling transcritical flows, where phase boundaries dynamically evolve, presents even greater numerical challenges than modeling fully supercritical flows \cite{ma2017entropy}.

Over the past three decades, transcritical and supercritical injection and combustion have been studied extensively. However, prior works have predominantly focused on single-component systems with fixed critical properties, where exceeding the critical point permits a single-phase "dense-fluid" assumption \cite{yang2000modeling}. In multi-component mixtures, the mixture critical pressure can significantly exceed those of the individual species, making accurate determination of the local mixture critical point essential for realistic modeling.

Capturing these multi-component phenomena requires a thermodynamic framework based on vapor-liquid equilibrium (VLE) theory, which enforces mechanical, thermal, and chemical equilibrium between phases to determine phase fractions, densities, and compositions. Coupling real-fluid equations of state (EOS) with VLE models accounts for real-fluid behavior \cite{zhang2022multicomponent}; however, this incurs severe computational costs, often consuming up to 90\% of total CFD execution time \cite{zhang2023numerical}. Each real-fluid VLE calculation requires nested inner and outer iterative loops to solve phase separation, chemical equilibrium, and property estimations. Because these calculations rely on Newton iterations that are highly sensitive to initial guesses, multi-component VLE routines are not only computationally expensive but also prone to non-convergence, frequently causing CFD simulation crashes \cite{yang2020real}.

Several strategies have been proposed to mitigate the computational burden of online real-fluid VLE calculations. Fathi~\textit{et al.}~\cite{fathi2021rapid} introduced a reduced-order VLE calculator by simplifying the multi-component interaction matrix, achieving moderate speedup but failing to guarantee numerical robustness. Alternatively, Zhang and Yang~\cite{zhang2023numerical} integrated \textit{in situ} adaptive tabulation (ISAT), demonstrating a 400-fold speedup; however, ISAT still relies on direct VLE solves to build its table, leaving it vulnerable to convergence failures \cite{srinivasan2026neural}.

Crucially, trained ANNs can interpolate across regions where direct VLE algorithms fail to converge, \textcolor{\mycolor}{thereby avoiding convergence difficulties associated with online iterative VLE calculations}.

To eliminate online VLE solves entirely, pre-generated VLE lookup tables can be used. However, multi-component pre-tabulation suffers from the "curse of dimensionality" - a table for just four components can easily exceed several TB, surpassing the random-access memory (RAM) capacity of state-of-the-art high-performance computing (HPC) nodes. Dynamic tabulation methods like ISAT also face table size limits; while ISAT can delete inactive leaves to bound memory usage and shorten search times, dropping leaves sacrifices domain coverage and triggers costly table re-expansions. Machine learning (ML) techniques, such as artificial neural networks (ANNs), overcome this memory constraint by performing non-linear regression with compact memory footprints (megabytes). Crucially, trained ANNs can interpolate across regions where direct VLE algorithms fail to converge, \textcolor{\mycolor}{thereby avoiding convergence difficulties associated with online iterative VLE calculations}.

Recent studies demonstrate the promise of ANN-based VLE modeling. Yue~\textit{et al.}~\cite{yue2023artificial} applied ANNs to fit experimental VLE data for diesel sprays, and Srinivasan~\textit{et al.}~\cite{srinivasan2024artificial} demonstrated that ANN-assisted VLE modeling in CFD simulations dramatically improves both numerical robustness and computational efficiency.

However, the efficiency gains of ANNs depend heavily on network size. In CFD applications where inferences occur at every grid cell and time step, large networks introduce substantial computational overhead. High-speed flows with extreme thermodynamic variations, such as hypersonic multiphase flows spanning $300\text{-}2000~\text{K}$ and $1\text{-}300~\text{bar}$, require larger network architectures to capture sharp property gradients near critical points without underfitting. Consequently, while deep ANNs \textcolor{\mycolor}{can improve predictive capability}, their evaluation cost can diminish overall computational efficiency.

To combine the robustness of ANNs with the evaluation speed of tabulation, this study proposes a novel hybrid ISAT-ANN framework to efficiently integrate Vapor-Liquid Equilibrium into Computational Fluid Dynamics simulations of transcritical and supercritical flows. While offline pre-tabulation is constrained by high dimensionality, ISAT dynamically tabulates ANN predictions during runtime \cite{pope1997computationally}. This hybrid approach leverages ANNs \textcolor{\mycolor}{to avoid iterative online VLE calculations while maintaining a compact representation}, while ISAT accelerates evaluation through rapid table lookups.

In parallel CFD simulations using the Message Passing Interface (MPI), localized ISAT tables on individual ranks can cause severe load imbalances as thermodynamic trajectories diverge across the physical domain. Although standalone neural networks feature a constant compute cost per grid cell that reduces the need for dynamic load balancing, ISAT lookups and table expansions create non-uniform execution times across ranks. To mitigate these imbalances, we implement a Dynamic Load Balancing (DLB) strategy that redistributes computational tasks across MPI ranks. Furthermore, the proposed ISAT-ANN and DLB framework is extended to support Adaptive Mesh Refinement (AMR), maintaining high computational efficiency in multi-scale flows with localized high-resolution features.

The remainder of this paper is organized as follows: Section~\ref{numerical_methods} details the numerical framework, including VLE theory, the ANN-assisted model, ISAT integration, DLB, and AMR coupling. Section~\ref{results} validates the ISAT-ANN framework on a Mach 5 shock-droplet interaction problem relevant to Rotating Detonation Engines (RDEs) and evaluates its computational performance, DLB efficacy, and AMR scalability. Finally, Section~\ref{conclusions} summarizes the primary findings and conclusions.

\section{\label{numerical_methods}Numerical Model and Computational Methods}
\subsection{Vapor Liquid Equilibrium} \label{VLE}
Vapor-Liquid Equilibrium (VLE) theory provides a framework for predicting phase change and phase separation by enforcing thermodynamic equilibrium for a specific state. These calculations are frequently referred to as "flash" problems, as they assume the instantaneous establishment of equilibrium under the defined constraints. The solution to the VLE problem is obtained by satisfying the thermodynamic equilibrium conditions established in Eq. \ref{eq:thermEq}.

\begin{align}
P_l=P_v=P,\quad T_l=T_v=T,\quad f_{i,l}=f_{i,v},\quad i=1,...,N. \label{eq:thermEq}
\end{align}
In Eq. \ref{eq:thermEq}, $P$ refers to the pressure, $T$ refers to the temperature, and $f_i$ is the fugacity of the $i^{th}$ species. Subscripts $l$ and $v$ represent the liquid and vapor phases, respectively. The species fugacity $f$ is a representation of the chemical potential of the species (eq. \ref{eq:fugacity}) and can be thought of as the real gas equivalent of partial pressure. 
\begin{align}
    f \stackrel{\text{def}}{=} p ^{\circ}  exp\left[\frac{\mu - \mu^{\circ}}{R T}\right]
    \label{eq:fugacity}
\end{align}
Thermodynamic equilibrium as described in Eq. \ref{eq:thermEq} imposes mechanical (pressure), thermal (temperature), and chemical (fugacity) equality between the liquid and vapor phases.

The equation of state (EoS) plays a vital role in the closure of the compressible CFD system of equations, as it provides a pathway to relate the thermodynamic variables (temperature, pressure, density, etc.). As the thermodynamics are governed by the solution of the VLE equations, the equation of state has been incorporated into the solver by deriving the fugacity formulation for the chosen EoS. The EoS is also used to calculate the required properties of both phases using appropriate compositions and mixing rules during the VLE solution iterations. The Peng-Robinson \cite{peng1976new} has been chosen as the EoS for all the studies discussed in this work. The PR EOS is formulated as shown in Eq. \ref{eq:preos}.
\begin{align}
    P=\frac{RT}{V-b}-\frac{a}{V\left(V+b\right)+b\left(V-b\right)}, 
    \label{eq:preos}
\end{align}
where $R$ is the ideal gas constant, $V$ is molar volume. The energy parameter $a$ and co-volume parameter $b$ for pure component $i$ are formulated as shown in equations \ref{eq:a} - \ref{eq:kappa}.
\begin{align}
    a_i &= 0.45724\frac{R^2 T_{c_i}^2}{P_{c_i}} \hat{a} \label{eq:a}, \\
    b_i&=0.07780\frac{RT_{c_i}}{P_{c_i}} \label{eq:b},\\
    \hat{a}&=\left[1+\kappa\left(1-\sqrt{\frac{T}{T_{c_i}}}\right)\right]^2,  \label{eq:ahat}\\
    \kappa&=0.37464+1.54226\omega_i-0.26992\omega_i^2 \label{eq:kappa}
\end{align}
where $T_{c_i}$ is the critical temperature of component $i$, $P_{c_i}$ is critical pressure of component $i$, and $\omega_i$ is the acentric factor of component $i$. These species properties are obtained from the NIST REFPROP database~\cite{LEMMON-RP10}. When the PR EOS is used for a multi-component mixture, mixing rules are used to calculate the energy parameter $a$ and co-volume parameter $b$~\citep{reid1977properties} as shown in Eqs. \ref{eq:mix1} and \ref{eq:mix2}.

\begin{align}
a&=\sum_i\sum_j\chi_i\chi_j\sqrt{a_ia_j}(1-k_{ij}),\label{eq:mix1}\\
b&=\sum_i\chi_i b_i,\label{eq:mix2}
\end{align}

The solution to the VLE problem for a specified global mixture and a pair of independent thermodynamic variables-such as temperature-pressure ($TP$), enthalpy-pressure ($HP$), or internal energy-volume ($UV$)-provides the necessary data regarding phase compositions and supplementary thermodynamic properties, including density and entropy. The selection of these state variables dictates the specific VLE solver architecture required, which in turn determines the computational complexity of the calculation.

\begin{figure}[H]
     \centering
     \begin{subfigure}[b]{0.4\textwidth}
         \centering
         \includegraphics[width=\textwidth]{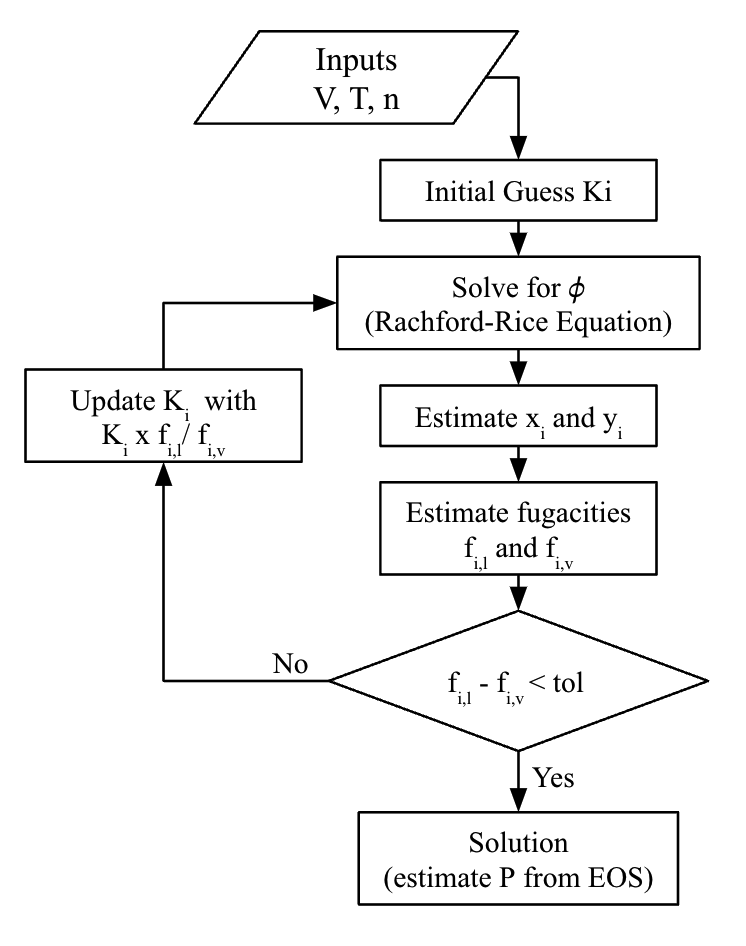}
         \caption{Flow chart of the VTn flash solver.}
         \label{fig:VTn}
     \end{subfigure}
     \hfill
     \begin{subfigure}[b]{0.52\textwidth}
         \centering
         \includegraphics[width=\textwidth]{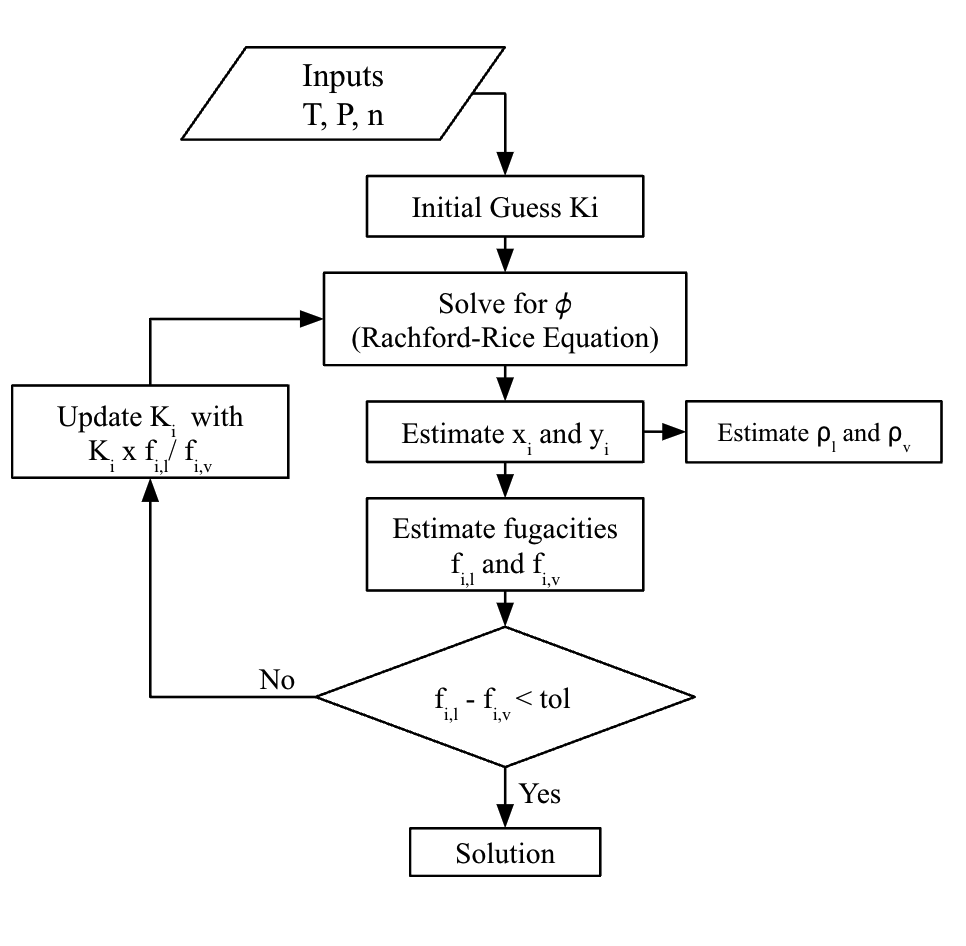}
         \caption{Flow chart of the TPn flash solver.}
         \label{fig:TPn}
     \end{subfigure}
     \hfill
        \caption{Temperature-based flash solvers.}
        \label{tempFlash}
\end{figure}

When temperature is utilized as a primary independent variable, a temperature-based flash algorithm (Fig. \ref{tempFlash}) is employed, with the specific solver contingent upon the identity of the second variable. If pressure is prescribed, the $TPn$ flash loop detailed in Fig. \ref{fig:TPn} is executed. Alternatively, if specific volume is the secondary variable, the $VTn$ flash solver (Fig. \ref{fig:VTn}) is invoked.

Solving either of the flash loops involves solving the Rachford-Rice equation (eq. \ref{eq:RR}), which is derived using the mass conservation law \cite{rachford1952procedure}:
\begin{align}
&\sum_{i=1}^{N}\bigg\{z_i\left(1-K_i\right)\bigg/\left[1+\left(K_i-1\right)\phi\right]\bigg\}=0, \label{eq:RR} \\
&\sum_{i=1}^{N}x_i=\sum_{i=1}^{N}y_i=1,  \label{eq:sumX} \\
&f_{i, l}\big/f_{i, g}=1,  \label{eq:fugequal}
\end{align}

where $x_i$ is the mole fraction of component $i$ in liquid phase, $y_i$ is the mole fraction of component $i$ in gas/vapor phase, $z_i$ is the mole fraction of component $i$ in the two-phase mixture (global), $\phi$ is the vapor mole fraction, $K_i$ is the equilibrium constant of component $i$ and $K_i=y_i/x_i$.
The equilibrium constants $K_i$ require an initial guess in the first iteration of the flash loop. The Wilson equation~\citep{wilson1964vapor} is commonly used to obtain the initial guess:
\begin{equation}
K_{i,wil}=e^{5.373(1+\omega_i)(1-T_{c_i}/T)}P_{c_i}/P. \label{wilsonEq}
\end{equation}

Eq. \ref{eq:RR} is solved using a Newton iteration to obtain the vapor fraction $\phi$, which is then used to estimate liquid and vapor phase species fugacities, which are used to update the equilibrium constant $K_i$ until the condition defined in eq. \ref{eq:fugequal} is satisfied. Further information about the solution to the flash loops can be found in \cite{ISAT_JCP, srinivasan2024artificial}.

\begin{figure}[h]
     \centering
     \begin{subfigure}[b]{0.45\textwidth}
         \centering
         \includegraphics[width=\textwidth]{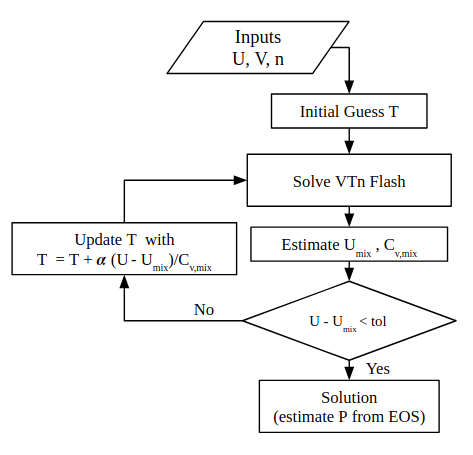}
         \caption{Flow chart of the UVn flash solver.}
         \label{fig:UVn}
     \end{subfigure}
     \hfill
     \begin{subfigure}[b]{0.45\textwidth}
         \centering
         \includegraphics[width=\textwidth]{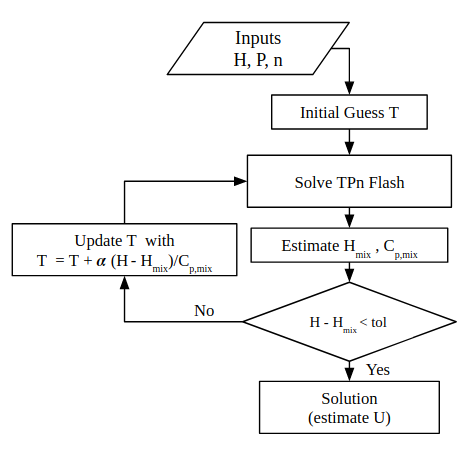}
         \caption{Flow chart of the HPn flash solver.}
         \label{fig:HPn}
     \end{subfigure}
     \hfill
        \caption{Energy-based flash solvers.}
        \label{energyFlash}
\end{figure}

Although temperature-based flash solvers offer superior reliability and ease of implementation, their direct application in real-fluid CFD is restricted by the non-linear relationship between internal energy or enthalpy and temperature. Under the ideal gas assumption, energy and enthalpy are functions of temperature alone; however, this assumption is invalid for real gases, where these properties also depend on pressure or specific volume. Furthermore, because the governing conservation equations are formulated in terms of energy or enthalpy rather than temperature, employing these properties as primary thermodynamic variables is more consistent with the numerical framework.

When internal energy and specific volume are selected as primary variables, the $UVn$ flash algorithm (Fig. \ref{fig:UVn}) is utilized; similarly, the $HPn$ flash (Fig. \ref{fig:HPn}) is executed when enthalpy and pressure are prescribed. As illustrated in the respective flowcharts, these energy-based solvers function by iteratively invoking their temperature-based counterparts, which introduces additional complexity into the solution procedure. This nested architecture can reduce solver reliability and significantly increase computational overhead. The substantial costs associated with energy-based VLE solvers are documented in \cite{ISAT_JCP, srinivasan2024artificial}, where VLE calculations have been shown to account for up to 90\% of the total CFD execution time.

\subsection{Artificial Neural Network (ANN) for VLE} \label{ANN}

Artificial Neural Networks (ANNs) have been deployed across a wide range of scientific disciplines and have recently been adapted for VLE-based CFD applications \cite{srinivasan2024artificial}. The primary motivation for integrating ANNs within CFD frameworks is to replace computationally intensive and potentially unreliable numerical solvers with a high-performance, easily implemented alternative. Conceptually, an ANN can be viewed as a high-fidelity compressed lookup table, where the training and testing errors represent the interpolative error associated with a given query. ANNs are particularly advantageous for VLE problems because they circumvent the "curse of dimensionality" inherent in traditional tabulation methods. For instance, conventional tables for four-component systems can require approximately 4 TB of storage \cite{ISAT_JCP}, exceeding the memory capacity of most High-Performance Computing (HPC) nodes. Consequently, ANNs offer a \textcolor{\mycolor}{non-iterative}, low-overhead alternative for incorporating VLE functionality into compressible CFD solvers without the prohibitive memory requirements of physical tables.

\begin{figure}[H]
    \centering
    \includegraphics[width=0.7\textwidth]{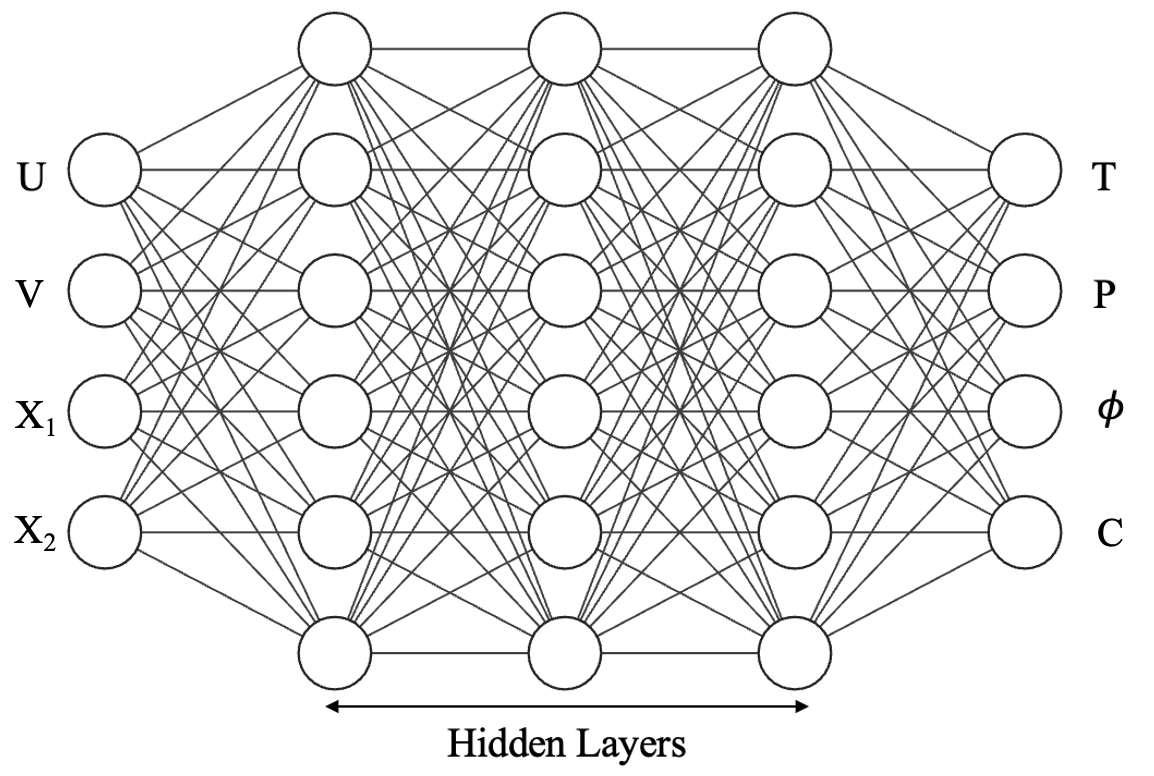}
    \caption{\textcolor{\mycolor}{Conceptual} ANN Structure for the UVn Flash Calculations.}
    \label{ANNstruct}
\end{figure}

The $UVn$ flash problem, being particularly well-suited for solvers utilizing conservative governing equations, is reformulated as a supervised learning problem using an Artificial Neural Network (ANN), as illustrated in Fig.~\ref{ANNstruct}. To ensure functional parity with the original numerical solver, the ANN architecture mirrors the inputs and outputs of the $UVn$ flash problem. \textcolor{\mycolor}{Specifically, in the load-balanced hybrid ISAT-ANN implementation, the four network inputs are the molar sensible internal energy ($u_m=eW_{\mathrm{mix}}$), density ($\rho$), and the species mass fractions ($Y_{\ce{C12H26}},Y_{\ce{N2}}$), supplied in the order $(u_m,\rho,Y_{\ce{C12H26}},Y_{\ce{N2}})$. Here, $e$ is the mass-specific sensible internal energy and $W_{\mathrm{mix}}$ is the mixture molecular weight.} In turn, it predicts the corresponding equilibrium state variables, including temperature ($T$), pressure ($P$), vapor fraction ($\phi$), and the speed of sound ($c$).

\textcolor{\mycolor}{The ANN employed in this study consists of nine hidden layers, each containing 64 neurons with Rectified Linear Unit (ReLU) activation functions. The architecture comprises an initial fully connected layer mapping the four inputs to 64 neurons, followed by eight applications of a shared 64-to-64 fully connected layer, and a final linear layer mapping to the four outputs. This configuration contains a total of 4,740 trainable parameters.} \textcolor{\mycolor}{The shared weights reduce the number of trainable parameters without eliminating the repeated computations associated with each hidden-layer evaluation. Although alternative network architectures were considered, a systematic architecture optimization study is not reported in the present work.}

The training of the ANN requires the generation of a comprehensive dataset, typically derived from a zero-dimensional (0-D) VLE framework. Although the ANN is designed to emulate the $UVn$ flash solver, the training data need not be generated strictly via a $UVn$ formulation. Instead, $TPn$ or $VTn$ flash codes can be utilized to solve the equilibrium problem, with the resulting phase compositions and vapor fractions used to calculate the necessary state variables, such as internal energy ($U$), specific volume ($V$), and the speed of sound ($c$). This methodology is fundamentally sound \textcolor{\mycolor}{provided that the two intensive properties are independent and the stable equilibrium branch is selected. At pure-fluid saturation, temperature and pressure alone do not determine the phase amount; an additional independent thermodynamic property is required to determine the vapor fraction. Near coexistence boundaries and critical states, the mapping may also become poorly conditioned.}

\textcolor{\mycolor}{For the present study, the primary thermodynamic dataset consists of 700,000 states generated using the $TPn$ flash formulation, covering temperatures from 270 to 2200 K and pressures from 0.3 to 20 MPa.}

This strategy offers two primary advantages. First, it yields a substantial reduction in computational costs during the data generation phase; by solving the $TPn$ flash, a single-loop algorithm, the system avoids the iterative overhead of the double-loop $UVn$ formulation. Second, this approach enhances numerical robustness, as the simpler $TPn$ structure \textcolor{\mycolor}{reduces} the convergence difficulties and complexities often encountered with double-loop algorithms in intricate VLE regimes. This increased stability facilitates the generation of the high-volume datasets required for CFD applications. Furthermore, utilizing $TPn$ for data generation simplifies the definition of input ranges, as temperature and pressure boundaries are more intuitively estimated than the specific internal energy and volume limits typically observed in high-speed flow simulations.

\textcolor{\mycolor}{The dataset is randomly divided into training, validation, and testing subsets, containing 72\%, 18\%, and 10\% of the available samples, respectively. Each input and output variable is normalized using the mean and standard deviation calculated from the training subset. These same normalization parameters are subsequently applied to the validation and testing subsets.}

\textcolor{\mycolor}{The ANN is trained using PyTorch \cite{paszke2017automatic} with the Adam optimization algorithm, an initial learning rate of $10^{-3}$, and a batch size of 256. The model is trained for 200 epochs using the mean absolute error (L1) loss function. A ReduceLROnPlateau learning-rate scheduler reduces the learning rate by a factor of 0.5 when the validation loss does not improve for four consecutive epochs. The model corresponding to the lowest validation loss is retained for deployment.}

\textcolor{\mycolor}{For the training dataset, the $TPn$ flash calculations employ a tangent-plane-distance (TPD) stability analysis to select the stable equilibrium branch, while states that do not converge to a valid equilibrium solution are excluded from the dataset.}

\begin{figure}[h]
    \centering
    \includegraphics[width=\textwidth]{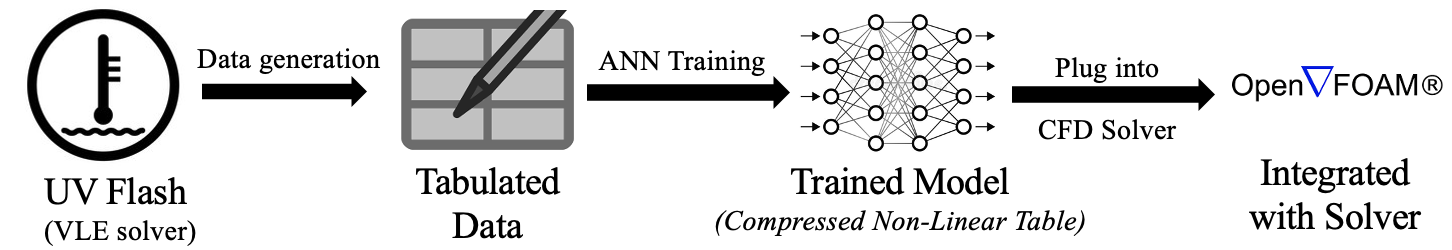}
    \caption{ANN aided VLE model training and coupling pipeline.}
    \label{Ann_pipe}
\end{figure}

The workflow involves converting the trained TensorFlow or PyTorch ANN model into the ONNX format within the Python environment, followed by seamless integration into the OpenFOAM solver using the ONNX C API. This ANN-aided VLE framework serves as a "plug-and-play" solution, enabling the conversion of standard multi-component CFD codes into VLE-capable solvers with minimal structural modifications. The comprehensive pipeline, extending from initial training to final solver coupling, is detailed in Fig. \ref{Ann_pipe}.

\subsection{In-Situ Adaptive Tabulation (ISAT) with ANN for VLE} \label{ISAT_ANN}

Architecturally, ISAT utilizes a binary tree-based data structure to facilitate efficient data retrieval. During runtime, the algorithm utilizes linear approximations \textcolor{\mycolor}{guided by user-defined error tolerances} to provide rapid estimates of the required properties, thereby bypassing full numerical calculations whenever \textcolor{\mycolor}{an input falls within an existing region of accuracy}. A critical factor in ISAT's efficacy is its \textcolor{\mycolor}{local error control mechanism, which evaluates interpolation errors during table growth to maintain the desired accuracy at sampled states}. Given these robust tabulation capabilities, ISAT is highly adaptable to the complexities of VLE problems, where it can be used to store and retrieve thermodynamic equilibrium states with high efficiency.

Every record in the ISAT table contains $(\mathbf{x}_0,\mathbf{y}_0,\left.\frac{\partial \mathbf{F}}{\partial \mathbf{x}}\right|_{\mathbf{x}_0},\textcolor{\mycolor}{\mathbf{G}})$, where $x_0$ is the input to the table, $y_0$ is the corresponding output, $\left.\frac{\partial \mathbf{F}}{\partial \mathbf{x}}\right|_{\mathbf{x}_0}$ is the Jacobian matrix used in the linear interpolation and $\textcolor{\mycolor}{\mathbf{G}}$ \textcolor{\mycolor}{is the stored matrix defining the region of accuracy (ROA)}.

The gradient (i.e., the Jacobian matrix $\left.\frac{\partial \mathbf{F}}{\partial \mathbf{x}}\right|_{\mathbf{x}_0}$) can be evaluated either numerically or analytically and is used for local linear approximation as shown in eq. \ref{eq:isat_inter}.
\begin{equation}
\mathbf{y}\approx\mathbf{y}_{linear}=\mathbf{y}_0+\left.\frac{\partial \mathbf{F}}{\partial \mathbf{x}}\right|_{\mathbf{x}_0}\cdot\left(\mathbf{x}-\mathbf{x}_0\right)
    \label{eq:isat_inter}
\end{equation}

The matrix $\mathbf{M}$ is used to define the region of accuracy (ROA), \textcolor{\mycolor}{which determines whether a new input can be evaluated using the stored linear approximation}. The ROA is defined by an inequality shown in eq. \ref{eq:RoA}.
\begin{equation}
  \left(\mathbf{x}-\mathbf{x}_0\right)^T \mathbf{M} \left(\mathbf{x}-\mathbf{x}_0\right) \textcolor{\mycolor}{\leq 1}
  \label{eq:RoA}
\end{equation}
The region satisfying this inequality is a hyper-ellipsoid, and hence the ROA is also called the ellipsoid of accuracy (EOA).

\textcolor{\mycolor}{To account for the different magnitudes of the input variables, the ISAT algorithm introduces a diagonal scaling matrix $\mathbf{S}$, such that the scaled input displacement is defined as $\mathbf{z}=\mathbf{S}^{-1}(\mathbf{x}-\mathbf{x}_0)$. The matrix $\mathbf{M}$ in Eq.~\ref{eq:RoA} is related to the stored EOA matrix $\mathbf{G}$ through $\mathbf{M}=\mathbf{S}^{-T}\mathbf{G}\mathbf{S}^{-1}$, resulting in the equivalent condition $\mathbf{z}^T\mathbf{G}\mathbf{z}\leq 1$. For the load-balanced ISAT-ANN model, the input variables consist of the two species mass fractions, molar sensible internal energy ($u_m=eW_{\mathrm{mix}}$), and density ($\rho$), where $e$ is the mass-specific sensible internal energy and $W_{\mathrm{mix}}$ is the mixture molecular weight. The input vector is therefore $\mathbf{x}=(Y_{\ce{C12H26}},Y_{\ce{N2}},u_m,\rho)^T$, with corresponding scaling factors $\mathbf{S}=\operatorname{diag}(1,1,10^3,10^2)$. The last two scaling factors have units of J/mol and kg/m$^3$, respectively. These scaling factors are specified independently of the mean and standard deviation normalization used for ANN inference.}

\textcolor{\mycolor}{When a new leaf is added to the ISAT table, its initial EOA is constructed using the local Jacobian and the prescribed input and output tolerances. Defining $\mathbf{J}=\left.\partial\mathbf{F}/\partial\mathbf{x}\right|_{\mathbf{x}_0}$, $\mathbf{R}=\operatorname{diag}(t_i^{-2})$, and $\mathbf{W}=\operatorname{diag}(\tau_j^{-2})$, where $t_i$ and $\tau_j$ represent the input and output tolerances, respectively, the initial EOA matrix is calculated as}

\begin{equation*}
\textcolor{\mycolor}{
\mathbf{G}_0 =
\mathbf{S}^T
\left(
\mathbf{R}+\mathbf{J}^T\mathbf{W}\mathbf{J}
\right)
\mathbf{S}.
}
\end{equation*}

\textcolor{\mycolor}{The input tolerances are $\mathbf{t}=(10^{-3},10^{-3},0.1,0.01)$, with the last two values expressed in J/mol and kg/m$^3$, respectively. The output tolerances for temperature, pressure, vapor fraction, and speed of sound are $\boldsymbol{\tau}=(2\,\mathrm{K},5\times10^4\,\mathrm{Pa},0.05,10\,\mathrm{m/s})$. The positive diagonal contribution from $\mathbf{R}$ ensures that the initial EOA matrix is symmetric positive definite, $\mathbf{M}=\mathbf{M}^T\succ 0$, for finite positive tolerances and nonzero scaling factors in exact arithmetic. This formulation also allows the EOA to be initialized when the Jacobian is rank deficient, without requiring matrix inversion or singular-value decomposition. Furthermore, both species mass fractions are retained in the ISAT input vector despite their sum-to-one constraint, without eliminating either variable. No additional singularity correction is applied during subsequent EOA growth.}

The ISAT table is organized as a binary tree rather than a conventional structured grid. Each internal node within this tree stores a vector that defines a cutting hyperplane, which partitions the local parameter space into two distinct half-spaces. The data subsets corresponding to these half-spaces are then managed by the respective sub-trees. Through this recursive hierarchical partitioning, the entire parameter space is subdivided into numerous localized cells, with the actual data records stored exclusively at the leaf nodes. During a search operation, the algorithm traverses each layer of the tree by evaluating which half-space contains the input parameters. This traversal continues until a leaf node is reached, allowing for high-speed data retrieval that scales logarithmically with the number of stored records.

During CFD runtime, the In-Situ Adaptive Tabulation (ISAT) algorithm facilitates three primary runtime operations: retrieval, growth, and addition. A retrieval is executed when an incoming input state is mapped to an existing leaf node and satisfies the defined Region of Accuracy (ROA). In this scenario, the algorithm returns a solution via linear interpolation using the stored Jacobian, thereby bypassing the full numerical solver. If the input state falls outside the current ROA of its associated leaf, ISAT initiates either growth or addition. First, \textcolor{\mycolor}{a reference evaluation is performed for the new input state, using ANN inference in the hybrid ISAT-ANN model or a direct VLE calculation in the conventional ISAT model}. This result is then compared against the interpolated prediction from the nearest leaf, followed by either the growth or addition step. The growth step is initiated if the interpolated prediction remains within a prescribed tolerance and the ROA of the existing leaf is expanded to encompass the new point. Addition is initiated if the prediction exceeds the error tolerance, \textcolor{\mycolor}{or if the maximum allowable input displacement is exceeded,} which results in the generation of a new leaf node. This involves partitioning the parameter space by inserting a new hyperplane between the existing leaf and the new data point.

\textcolor{\mycolor}{To determine whether the EOA can be expanded, the interpolation error is calculated by comparing the linear approximation against the reference evaluation at the new input state. The error is normalized using the prescribed output tolerances, and growth is permitted when}

\begin{equation*}
\textcolor{\mycolor}{
\epsilon =
\left[
\sum_j
\left(
\frac{y_{\mathrm{linear},j}-y_{\mathrm{ref},j}}
{\tau_j}
\right)^2
\right]^{1/2}
\leq 1.
}
\end{equation*}

\textcolor{\mycolor}{Here, $\mathbf{y}_{\mathrm{ref}}$ represents the ANN prediction for the hybrid ISAT-ANN model. When the error satisfies this criterion, the EOA is expanded using a rank-one update of the stored matrix, given by}
\begin{equation*}
\textcolor{\mycolor}{
\mathbf{G}_{\mathrm{new}} =
\mathbf{G}+
\frac{1-q}{q^2}
\mathbf{G}\mathbf{w}\mathbf{w}^T\mathbf{G},
\qquad
q=\mathbf{w}^T\mathbf{G}\mathbf{w},
}
\end{equation*}
\textcolor{\mycolor}{where $\mathbf{w}=1.001\mathbf{S}^{-1}(\mathbf{x}-\mathbf{x}_0)$. The additional factor of 1.001 slightly extends the growth beyond the evaluated input state. This procedure allows the EOA to expand as new thermodynamic states are encountered, reducing the number of direct evaluations required during subsequent time steps. However, the interpolation error is checked only at the sampled input during growth, while states retrieved from an existing EOA are not independently evaluated against the reference function. Consequently, the prescribed tolerance provides local error control relative to the ANN predictions, rather than a guarantee that the interpolation error remains below the tolerance throughout the entire EOA. Furthermore, this criterion does not account for the underlying approximation error of the ANN relative to the direct VLE solution.}

ISAT has been successfully adapted for VLE-based CFD, where the table is populated dynamically during runtime via direct evaluation of the governing thermodynamic equations \cite{ISAT_JCP}. This approach has demonstrated substantial speedups, rendering multicomponent real-fluid simulations computationally viable.

Despite these advancements, ISAT-aided VLE frameworks still encounter reliability challenges. These issues stem from the reliance on direct evaluation; the iterative VLE solvers often struggle with convergence or divergence when subjected to the extreme gradients and thermodynamic states typical of complex cases, such as high-speed (Mach 5) shock-droplet interactions.

To address the reliability constraints, an artificial neural network (ANN) based VLE model was developed (described above), which replaced the direct evaluation of the VLE equations and provided more significant speedups than the ISAT-only approach (while using direct evaluation) for test cases discussed in \cite{srinivasan2024artificial}. The speedups provided were directly dependent on the size of the ANN employed, and larger ANNs required larger inference times.

However, running complex thermophysical simulations such as high-speed flows with multiphase systems requires additional considerations. To address the challenges posed by large thermophysical variations, such as those encountered in Mach 5 shock-droplet interactions relevant to Rotating Detonation Engine (RDE) conditions, the VLE framework has been further evolved. In these high-speed regimes, a standard ANN requires a significantly larger architecture-often 10 to 20 times the size of models used for lower Mach numbers \cite{srinivasan2024artificial}-to meet the stringent training criteria (L1 regularization error < 0.1\%). This increased network depth and width result in a non-negligible rise in computational latency during CFD execution.

To optimize performance under these conditions, the existing ISAT-aided VLE infrastructure was refined to create a hybrid ISAT-ANN VLE model. In this configuration, the ANN is utilized to replace the traditional direct evaluation step \textcolor{\mycolor}{during table misses, including growth checks and the addition of new leaves}. Furthermore, to maintain the linear approximation capabilities within the ISAT leaves, a numerical Jacobian is computed to substitute the analytical Jacobian utilized in previous iterations. This hybrid approach leverages the speed of ISAT's retrieval logic while employing the ANN as a robust, non-iterative alternative to fragile numerical VLE solvers.

\begin{figure}[h]
    \centering
    \includegraphics[width=0.75\textwidth]{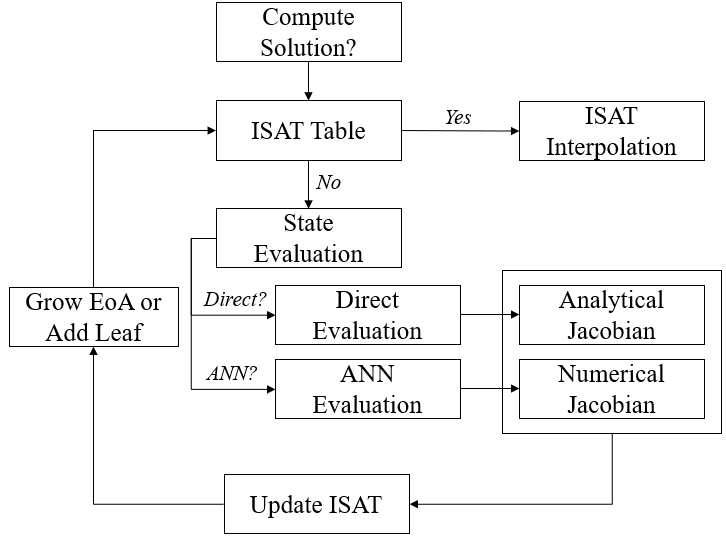}
    \caption{In-Situ Adaptive Tabulation (ISAT) for Vapor Liquid Equilibrium (VLE) based CFD. Both ANN and direct evaluation options are available to be used along with ISAT.}
    \label{fig:ISAT_ANN}
\end{figure}

Figure \ref{fig:ISAT_ANN} illustrates the integration of ANN-based inference into the ISAT pipeline, while \textcolor{\mycolor}{also showing direct numerical evaluation as an alternative operating mode}. To ensure maximum efficiency during the CFD runtime, the ONNX runtime is utilized for model inference, as it represents the most performant option for high-frequency calls within a C++ environment.

As ONNX runtime optimizations prioritize execution speed over backpropagation capabilities, it does not support automatic gradient evaluation. Consequently, the Jacobians required for the ISAT linear approximations must be estimated numerically. These numerical Jacobians are computed via a \textcolor{\mycolor}{forward finite-difference scheme, using fixed perturbations of 0.01 for each species mass fraction, 1 J/mol for molar sensible internal energy, and 1 kg/m$^3$ for density. The species mass fractions are perturbed individually without enforcing their sum-to-one constraint, while derivatives are evaluated for all four output variables, including the speed of sound}. This approach \textcolor{\mycolor}{provides the local gradients required for linear interpolation, with the resulting approximation accuracy assessed through the error criterion described above}.

\subsection{Adaptive Mesh Refinement (AMR)} \label{AMR}

The computational overhead introduced by VLE solutions represents a critical bottleneck in large-scale CFD simulations, a challenge that is further exacerbated by stringent mesh resolution requirements. Capturing fine-scale physics-such as the growth of interfacial instabilities on a droplet subjected to high-speed flow-demands a spatial resolution capable of resolving features significantly smaller than the primary flow scales. For transient phenomena like Mach 5 shock-droplet interactions, standard practice suggests 100-500 grid points across the droplet diameter. Employing a uniform mesh to meet these requirements throughout the entire domain results in cell counts 4-8 times higher than necessary, leading to prohibitively slow and expensive simulations.

Adaptive Mesh Refinement (AMR) offers a more efficient alternative to static, uniform grids for such transient problems by dynamically modifying the mesh topology in response to evolving flow features. Due to its hierarchical and adaptive nature, AMR minimizes computational expenditure by selectively refining only the regions of interest-such as shock fronts and material interfaces-while maintaining a coarser base resolution elsewhere. This capability allows researchers to capture high-fidelity physical details without the need for manual generation of complex, multi-block structured meshes, effectively balancing the competing demands of local precision and global efficiency.

OpenFOAM \cite{of6} provides an inbuilt AMR feature called dynamic mesh, which allows for modifying the mesh based on user-selected fields and thresholds. This feature also allows defining the number of buffer layers to be introduced between different mesh levels and the criteria to coarsen the mesh back to its base resolution. Users can also choose the frequency of mesh update (either refine or coarsen) based on the number of time steps. However, this feature works only with 3D hexahedral cells \cite{of6,baniabedalruhman2015dynamic} and cannot be used in 2D cases.

\begin{figure}[h]
    \centering
    \includegraphics[width=\textwidth]{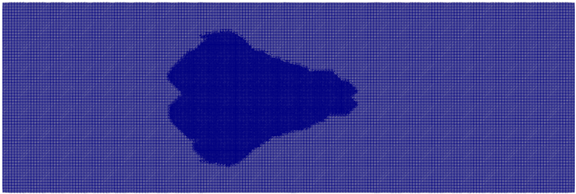}
    \caption{Adaptive mesh refinement enabled mesh visualized for a 2D shock-droplet interaction case to resolve droplet dynamics.}
    \label{fig:AMR}
\end{figure}

Given that 2D simulations offer valuable physical insights, serve as essential benchmarks for model validation, and are significantly more cost-effective than 3D counterparts, the implementation of a 2D AMR capability is crucial for VLE-based CFD research. To address this, the native 3D AMR framework in OpenFOAM was adapted for 2D applications by Sun et al. \cite{sun2023detonationfoam}, utilizing the methodology originally proposed by Baniabedalruhman \cite{baniabedalruhman2015dynamic}. This adaptation extends AMR functionality to both 2D Cartesian and axisymmetric meshes. Comprehensive details regarding the implementation and library architecture are provided in \cite{sun2023detonationfoam}. Figure \ref{fig:AMR} illustrates the application of this adapted 2D AMR library to a shock-droplet interaction case.

To evaluate the performance of the developed numerical models under high-speed conditions, a Mach 5 shock-droplet interaction was selected as the benchmark case. In this scenario, AMR is employed to resolve the droplet interface and the immediate surrounding flow field. The mass fraction of the liquid chemical species (C12H26) serves as the refinement criterion; the algorithm selectively resolves regions where the mass fraction values range between 0.001 and 0.999. Conversely, the mesh is coarsened when values fall outside these thresholds, ensuring computational resources are focused on the phase boundaries. The refinement frequency is established at every 10 time steps, with \textcolor{\mycolor}{1-4 buffer layers} implemented between successive refinement levels to maintain numerical stability and smooth transitions across the mesh hierarchy. \textcolor{\mycolor}{A maximum refinement level of 2 is employed.}

\subsection{Dynamic Load Balancing (DLB) for VLE} \label{DLB_VLE}

In parallel CFD simulations, particularly those employing the Message Passing Interface (MPI) for inter-process communication, ISAT tables typically remain private to each individual MPI rank. This localization restricts the global potential for computational cost reduction, as redundant thermodynamic states must be independently tabulated by each process. To maximize the efficiency of sophisticated hybrid models-such as the combined ISAT-ANN framework-it is imperative to maintain performance scalability in parallel environments.

Furthermore, the overall execution speed of a parallel CFD simulation is often throttled by a single MPI rank due to an uneven distribution of computational workloads across processors. To mitigate these bottlenecks, Dynamic Load Balancing (DLB) is implemented. In this context, "load balancing" refers to the nearly uniform distribution of total Central Processing Unit (CPU) requirements across all available MPI ranks, while "dynamic" signifies that this redistribution occurs during runtime in response to the instantaneous workload fluctuations encountered by each process.

To implement DLB within the current VLE framework, an existing algorithm originally developed for estimating chemical source terms in reacting flows \cite{tekgul2021dlbfoam} was adapted and further optimized. This enhanced DLB approach ensures that the thermodynamic overhead is equitably distributed, thereby improving the parallel efficiency of the multicomponent real-fluid solver.

\begin{figure}[htpb]
     \centering
     \begin{subfigure}{0.55\textwidth}
         \centering
         \includegraphics[width=\linewidth]{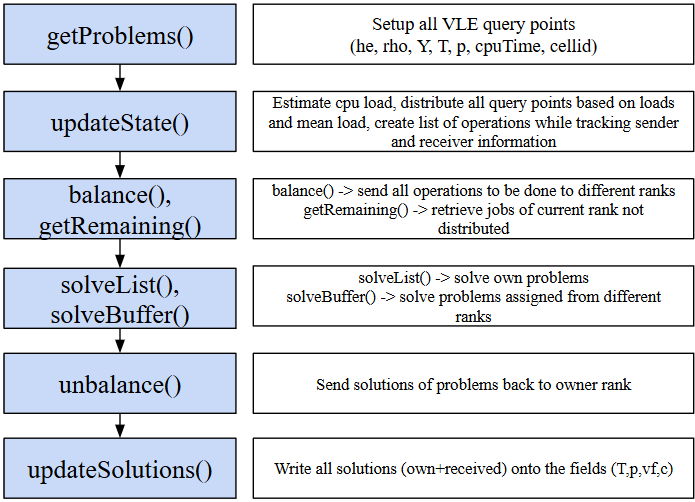}
         \caption{Functional calls of the DLB algorithm for VLE}
         \label{fig:DLB_diagram}
     \end{subfigure}
     \hfill
     \begin{subfigure}[b]{0.4\textwidth}
         \centering
         \includegraphics[width=\linewidth]{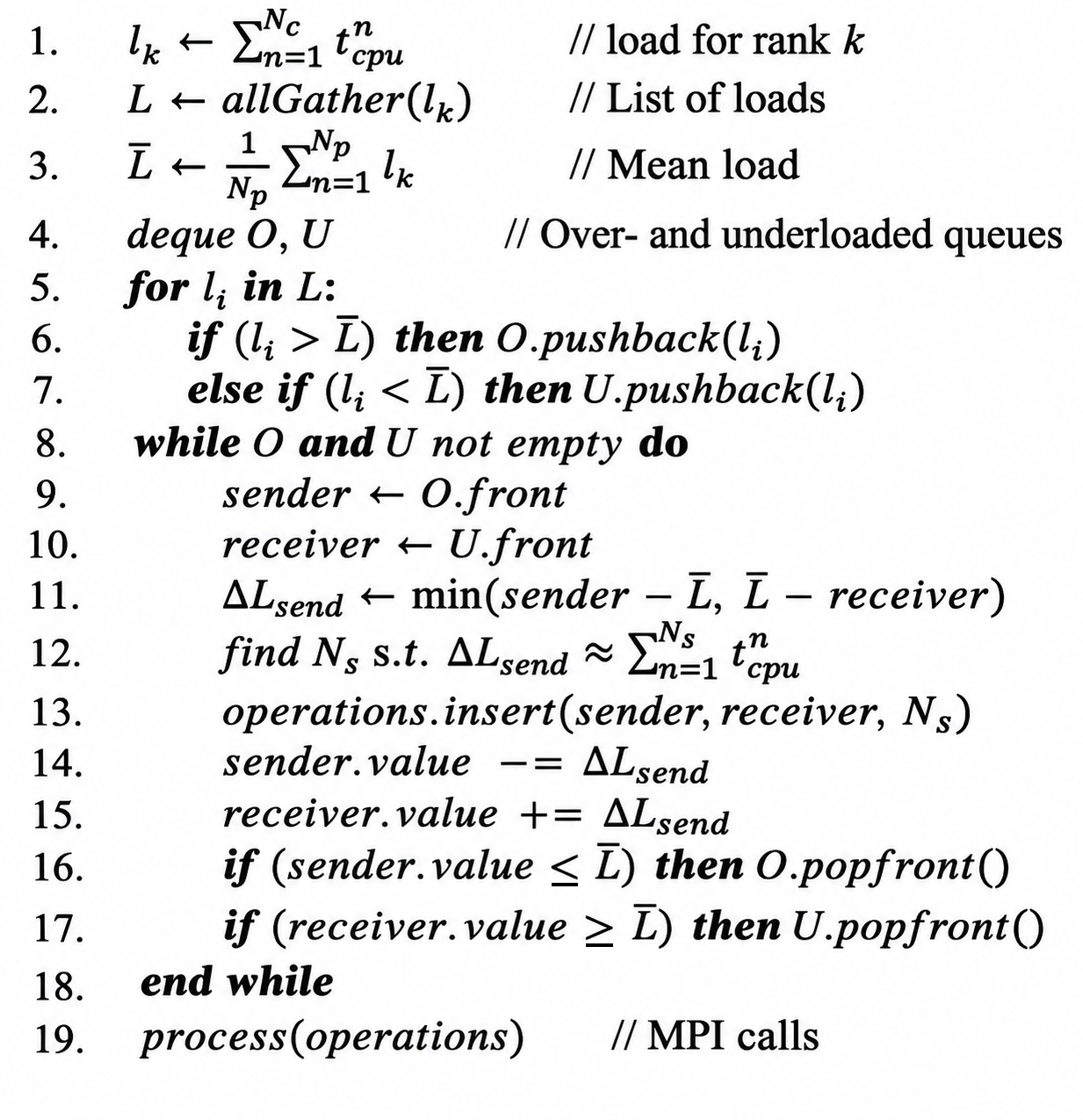}
         \caption{\textcolor{\mycolor}{Modified DLB algorithm with utilization of double-ended queues and greedy algorithm}}
         \label{fig:DLB_flowchart}
     \end{subfigure}
     \caption{Dynamic Load Balancing (DLB) for ISAT-ANN-based VLE solutions: (a) shows the flow of DLB through function calls; (b) describes the details of the modified DLB algorithm.}
     \label{Fig:DLB}
\end{figure}

Figure~\ref{Fig:DLB} illustrates the Dynamic Load Balancing (DLB) strategy tailored for ISAT-ANN aided VLE solutions, with the functional workflow detailed in Fig.~\ref{fig:DLB_diagram}. The process commences with each MPI rank independently assembling its local problem set by traversing its assigned subdomain to generate a list of VLE query points. Each query is encapsulated in a data structure comprising enthalpy, density, mass fractions, temperature, and pressure, alongside the specific cell ID and the recorded CPU time, defined as the cumulative computational duration required to solve the VLE system for that specific cell. While the cell ID is critical for mapping the solution back to the mesh, it is noteworthy that these IDs are rank-local and hold no intrinsic meaning to external MPI ranks.

Following problem assembly, the updateState function redistributes these queries across the available MPI ranks based on current CPU loads to ensure that every rank converges toward the global mean workload. The balance operation executes the necessary MPI send and receive calls to migrate problems to their target ranks, while the getRemaining function manages the internal assignment of problems that remain on their originating rank. These queries are aggregated into lists, which are subsequently processed by the solveList and solveBuffer routines using the ISAT-ANN VLE framework. Once solved, the unbalance call facilitates the return of processed data from receiver ranks back to the original sender ranks. Finally, the updateSolution routine utilizes the stored cell IDs to accurately populate the thermodynamic fields with the newly computed equilibrium values.

The Dynamic Load Balancing (DLB) algorithm was modified to optimize workload distribution through a greedy approach, eliminating unnecessary loops and adopting more efficient data structures to achieve speedups. In the original algorithm, load distribution requires sorting the complete load list $L$ using an $O(N \log N)$ sorting algorithm. It then applies a two-pointer technique, with one scanning from the start and the other from the end, to strictly match the most overloaded MPI rank with the most underloaded rank. While this guarantees exact pairing of extremes, the sorting phase creates significant computational overhead as the number of processes scales.

The modified algorithm circumvents this bottleneck by relaxing the strict rank pairing requirement. It first calculates an estimated mean load across all processes and uses an $O(N)$ linear loop to categorize ranks into two separate double-ended queues (deques): one for overloaded ranks and another for underloaded ranks. \textcolor{\mycolor}{Ranks whose workloads are exactly equal to the estimated mean are considered balanced and are therefore excluded from both queues, as no workload redistribution is required for these ranks.} Workloads are then migrated directly from any overloaded rank to any underloaded rank until either process reaches the mean load threshold. \textcolor{\mycolor}{Once a processor reaches the mean workload during redistribution, it is no longer considered for additional workload transfers in the current balancing step. The algorithm then continues redistributing work among the remaining overloaded and underloaded processors until no further eligible transfers remain.}

As illustrated in Fig.~\ref{fig:DLB_flowchart}, this updated implementation streamlines the load-rebalancing process by replacing strict rank pairing with a fast, queue-based allocation strategy, eliminating the $O(N \log N)$ sorting phase. As a result, the system achieves lower re-balancing latency and higher speedups while maintaining an equitable load distribution across all MPI processes.

\subsection{Compressible CFD Solver} \label{CFD_Solver}
The equations of mass, momentum, and energy conservation, together with component transport, can be written as:
\begin{equation}
\frac{\partial \rho}{\partial t}+\nabla \cdot(\rho u)=0;
\label{equ:mass}
\end{equation}

\begin{equation}
\frac{\partial \rho u}{\partial t}+\nabla \cdot(\rho u \otimes u)=-\nabla P+\nabla \cdot \tau;
\label{equ:momentum}
\end{equation}

\begin{equation}
\frac{\partial \rho E}{\partial t}
+\nabla \cdot(\rho E u)
=-\nabla \cdot(u P)
+\nabla \cdot(u \cdot \tau)
+\textcolor{\mycolor}{\nabla \cdot\left(
q-\sum_{s=1}^{n_s}h_s j_s
\right)};
\label{equ:energy}
\end{equation}

\begin{equation}
\frac{\partial \rho Y_{s}}{\partial t}+\nabla \cdot\left(\rho Y_{s} u\right)=-\nabla \cdot j_{s}
\label{equ:species},    \qquad \qquad  s=1,...,n_{s}-1,
\end{equation}

where $\rho$ is the mixture density, $P$ is the pressure, $\boldsymbol{u}$ is the velocity vector, $\tau$ is the viscous stress tensor, $\boldsymbol{q}$ is the heat flux, $E = e + \frac{1}{2} u \cdot u$ is the total energy. \textcolor{\mycolor}{The heat-flux vector is defined using the sign convention $\boldsymbol{q}=\lambda_{\mathrm{th}}\nabla T$, where $\lambda_{\mathrm{th}}$ denotes the thermal conductivity evaluated using the Chung model described subsequently. Accordingly, $\boldsymbol{q}$ is the negative of the conventional Fourier heat flux, $\boldsymbol{q}_{F}=-\lambda_{\mathrm{th}}\nabla T$, making the positive divergence term in Eq.~(18) consistent with the conductive contribution to the energy equation.} In Eq.~(\ref{equ:species}), $Y_{s}$ and $j_{s}$ are the mass fraction and diffusive mass flux of component $s$, respectively, while $n_{s}$ is the total number of components. \textcolor{\mycolor}{Species diffusion is represented in the component transport equations, while the corresponding diffusive species-enthalpy contribution is included in the total energy equation through $-\nabla\cdot\left(\sum_{s=1}^{n_s}h_s j_s\right)$, where $h_s$ is the specific enthalpy of component $s$. This contribution is treated separately from the conductive heat-flux vector $\boldsymbol{q}$.}

The CFD solver is based on the central-upwind scheme \cite{kurganov2001semidiscrete,greenshields2010implementation}, a fully conservative (FC) formulation implemented within the OpenFOAM environment \cite{of6}. During each temporal integration step, the conservative variables, comprising density ($\rho$), momentum ($\rho u$), total energy ($\rho E$), and species mass fraction ($\rho Y_s$), are updated according to the governing transport equations (Eqs. \ref{equ:mass}-\ref{equ:species}). To close the system of equations, the equation of state (EOS) is integrated directly into the Vapor-Liquid Equilibrium (VLE) flash calculation, which is solved using the introduced ISAT-ANN-aided VLE model at every time step to ensure thermodynamic consistency throughout the simulation. \textcolor{\mycolor}{For the shock-droplet simulations, the thermodynamic model employs the Peng-Robinson equation of state with temperature-dependent JANAF caloric properties, while the prescribed initial shock conditions are clarified separately in Section~3.1.}

The generalized method of Chung et al. \cite{chung1988generalized} is employed to evaluate the dynamic viscosity and thermal conductivity under high-pressure transcritical conditions. This approach provides accurate estimations for polar, non-polar, and associating pure fluids as well as their mixtures \cite{chung1988generalized}. Both dynamic viscosity and thermal conductivity are determined using a consistent functional form:
\begin{equation}
\lambda = \lambda_0 \lambda^* + \lambda_p, \label{transp}
\end{equation}
where $\lambda$ denotes the transport property of interest. In this formulation, $\lambda_0$ represents the gas-phase property at the low-pressure limit, while $\lambda^*$ and $\lambda_p$ serve as high-pressure corrections. The model is structured such that at elevated pressures, $\lambda_p$ becomes the dominant contributing term relative to the product $\lambda_0 \lambda^*$. Conversely, at low pressures, $\lambda^*$ approaches unity and $\lambda_p$ becomes negligible, allowing Eq. \ref{transp} to recover the dilute gas property $\lambda_0$. Consequently, the model provides a continuous and smooth transition across the pressure spectrum from the dilute gas to the dense fluid regime.

Mass diffusivity is characterized using a mixture-averaged diffusion model. Following the formulation proposed by Kee et al. \cite{kee1996chemkin}, the mass diffusion coefficient $D_i$ for the $i$-th component is defined as:
\begin{equation}
D_i=\frac{1-Y_i}{\sum^N_{j\neq i}X_j/D_{j,i}}, 
\label{massdiff}
\end{equation}
where $Y_i$ and $X_i$ denote the mass and mole fractions of component $i$, respectively. The binary diffusion coefficient, $D_{i,j}$, is determined using Fuller’s model \cite{fuller1966new}. To ensure accuracy under transcritical and high-pressure conditions, Takahashi’s correction \cite{takahashi1975preparation} is applied to the binary coefficients. This integrated approach accounts for both the molecular interactions in multi-component mixtures and the significant effects of high pressure on diffusive transport.

\section{Results and Discussions} \label{results}

\subsection{Domain and Configuration - Mach 5 Shock Droplet Interaction}
\begin{figure}[H]
    \centering
    \includegraphics[width=\textwidth]{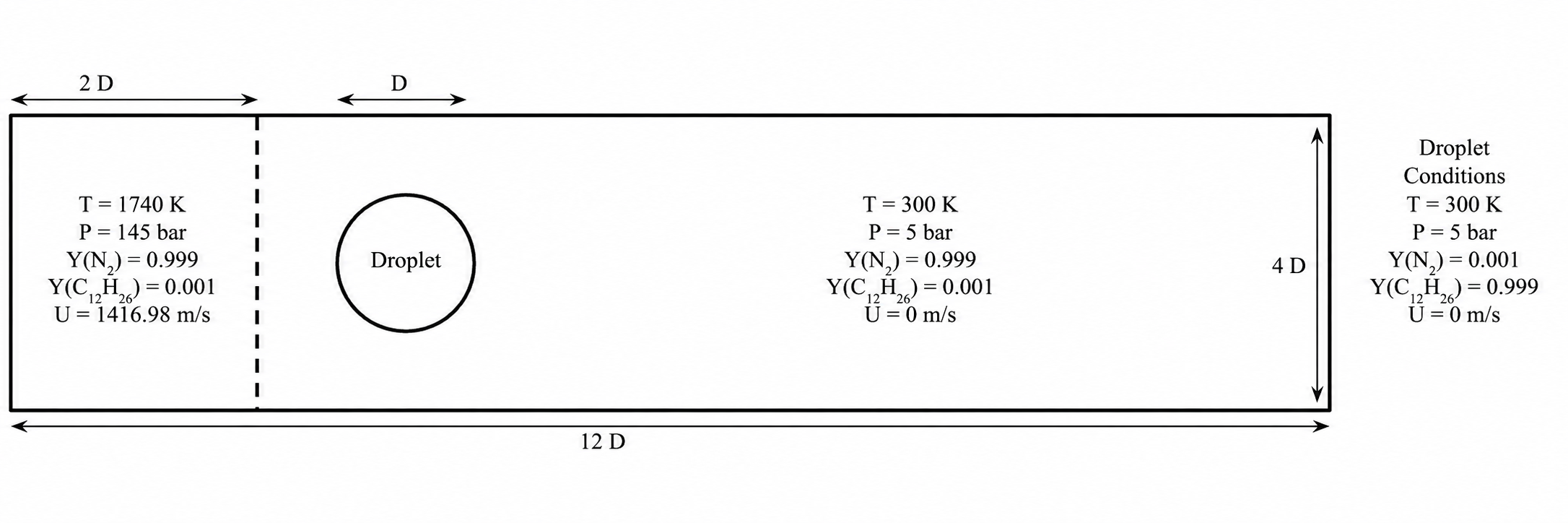}
    \caption{Domain and configuration of \textcolor{\mycolor}{the nominal} Mach 5 shock droplet interaction simulation used as a test case to evaluate model performance. Droplet diameter set to \textcolor{\mycolor}{$250 \mu m$} for all cases simulated.}
    \label{domain}
\end{figure}

Within a vast range of numerical tools described to facilitate the numerical simulations of VLE-based CFD simulations at high-pressure operating conditions, a suitable test case to evaluate the performance of the developed methods is required. The required test case is chosen to be a two-dimensional (2D) interaction between a shock wave and a liquid droplet at \textcolor{\mycolor}{nominal} Mach 5 conditions, relevant to multiphase detonation. The droplet experiences rapid thermodynamic changes as the shock passes through it and introduces the multi-component VLE effect, making it a challenging problem to simulate. The Mach 5 shock is initiated using a high-pressure and high-temperature discontinuity along with a post-shock velocity. \textcolor{\mycolor}{The upstream gas is initialized at $T_1=300$ K and $P_1=5$ bar with zero velocity, while the post-shock region is prescribed at $T_2=1740$ K, $P_2=145$ bar, and $u_2=1416.9752$ m/s. The pressure and temperature ratios, $P_2/P_1=29$ and $T_2/T_1=5.8$, coincide with the calorically perfect-gas normal-shock relations for Mach 5 with $\gamma=1.4$. However, these initial conditions are prescribed directly rather than obtained by solving the Rankine-Hugoniot relations using the real-fluid equation of state. During initialization, the Peng-Robinson equation of state, JANAF caloric model, and TP equilibrium flash are used to determine the corresponding density, internal energy, vapor fraction, and speed of sound. The shock propagation speed is not independently prescribed, and the initial discontinuity subsequently evolves under the conservative CFD formulation described in Section~\ref{CFD_Solver}.} The simulations use a rectangular domain with dimensions parameterized with respect to droplet diameter (D) as shown in Figure \ref{domain}, which is set to \textcolor{\mycolor}{$250 \mu m$}. The droplet is composed of \ce{C12H26}. The surrounding environment is filled with \ce{N2}. \textcolor{\mycolor}{The ambient gas composition is specified by mass fractions of $Y_{\mathrm{N_2}}=0.999$ and $Y_{\mathrm{C_{12}H_{26}}}=0.001$.} The domain configuration is shown in Fig.~\ref{domain}. 
\textcolor{\mycolor}{The computational domain extends from $x=-0.5$ to $2.5$ mm and from $y=0$ to $1$ mm. At the left boundary, the inlet velocity is prescribed as $\mathbf{u}=(1416.9752,0,0)$ m/s, while zero-gradient conditions are applied to pressure, temperature, and species mass fractions. At the outlet and the upper and lower boundaries, zero-gradient conditions are applied to all flow variables. The simulations are configured as two-dimensional cases with one cell in the third direction and empty boundary conditions on the front and back planes. These boundary conditions are maintained across all four mesh resolutions considered in the mesh-refinement study.}

\begin{figure}[htbp]
    \centering
    \includegraphics[width=\textwidth]{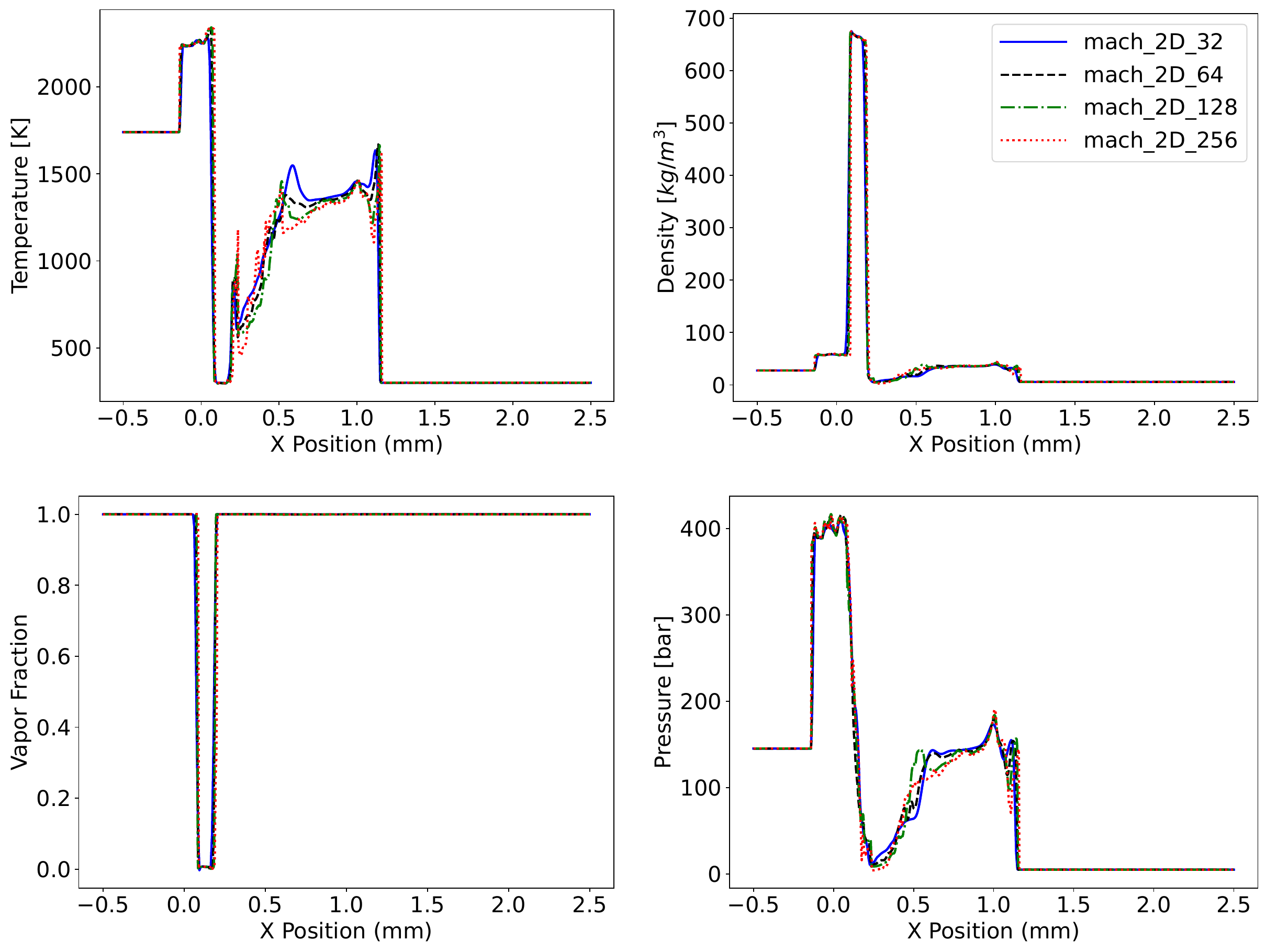}
    \caption{\textcolor{\mycolor}{Mesh-refinement study of the nominal Mach 5 shock-droplet interaction at $t=0.8~\mu$s. Centerline profiles at $y=0.5$ mm are compared for four uniform mesh resolutions with 32, 64, 128, and 256 cells across the initial droplet diameter: (a) temperature, (b) density, (c) vapor fraction, and (d) pressure.}}
    \label{fig:mesh_convergence}
\end{figure}

\textcolor{\mycolor}{To evaluate the influence of mesh resolution on the numerical solution, a mesh-refinement study was conducted using four static meshes with 32, 64, 128, and 256 cells across the initial droplet diameter of 250~$\mu$m. The corresponding mesh configurations are $384\times128$, $768\times256$, $1536\times512$, and $3072\times1024$ cells, with spatial resolutions of 7.8125, 3.90625, 1.953125, and 0.9765625~$\mu$m, respectively. All four cases utilize the same computational domain and prescribed initial conditions, without adaptive mesh refinement. The centerline distributions of temperature, density, vapor fraction, and pressure at $t=0.8~\mu$s are compared in Fig.~\ref{fig:mesh_convergence}.}

\textcolor{\mycolor}{As shown in Fig.~\ref{fig:mesh_convergence}(c), the vapor fraction profiles exhibit nearly identical distributions across all four mesh resolutions, indicating that the centerline two-phase region and the location of the liquid-vapor interface are already well resolved using the coarsest mesh with 32 cells across the droplet diameter. Similarly, the density profiles show close agreement throughout the domain. In contrast, the temperature and pressure distributions exhibit more noticeable differences downstream of the droplet. Comparison with the vapor fraction profiles indicates that these deviations occur predominantly in the single-phase region and hence are not due to real-fluid VLE modeling, while the two-phase region remains largely unaffected by mesh refinement.}

\textcolor{\mycolor}{These differences can be attributed to the development of hydrodynamic instabilities during the shock-droplet interaction. The acceleration of the droplet and the velocity gradients near its interface promote the development of interfacial instabilities, which subsequently influence the downstream flow. As the mesh resolution increases, smaller-scale disturbances and flow structures can be resolved, resulting in additional variations in the temperature and pressure distributions. This behavior is consistent with established instability mechanisms in shock-induced droplet deformation and breakup \cite{Sharma2021ShockAerobreakup,Rossano2024ScaleResolving}. In summary, 
the close agreement in vapor fraction demonstrates that the centerline phase distribution is adequately resolved even on the coarsest mesh. The remaining differences are primarily associated with the resolution of smaller-scale unstable flow structures rather than changes in the predicted location of the two-phase region.}

\subsection{Qualitative Analysis} \label{qualitative}

\begin{figure}[p]
    \centering

    \begin{subfigure}{\textwidth}
        \centering
        \includegraphics[width=\linewidth]{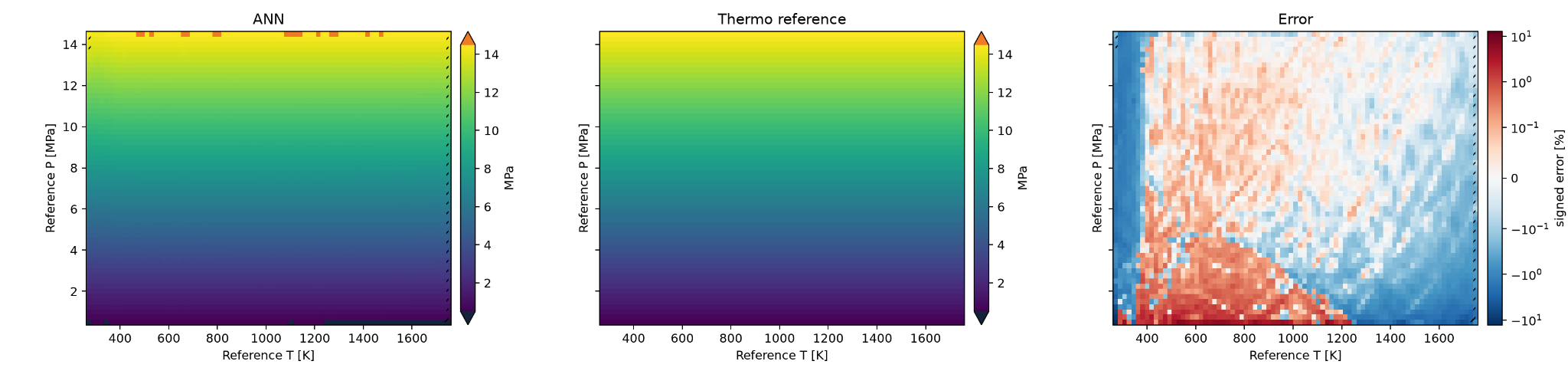}
        \caption{\textcolor{\mycolor}{Pressure}}
        \label{fig:VLE_pressure}
    \end{subfigure}

    \begin{subfigure}{\textwidth}
        \centering
        \includegraphics[width=\linewidth]{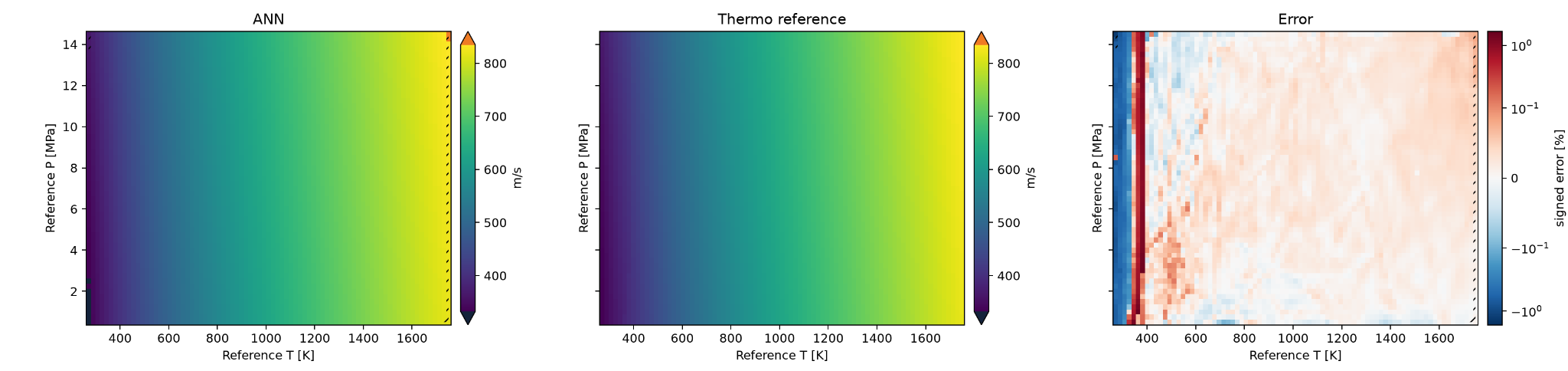}
        \caption{\textcolor{\mycolor}{Speed of sound}}
        \label{fig:VLE_sound}
    \end{subfigure}

    \caption{\textcolor{\mycolor}{Comparison of ANN predictions with reference VLE calculations for a nitrogen-dodecane mixture ($Y_{\mathrm{C_{12}H_{26}}}=0.01$, $Y_{\mathrm{N_2}}=0.99$) over temperatures of 270-1750 K and pressures of 0.5-14.5 MPa. Each row presents the ANN prediction, reference VLE calculation, and corresponding error. Hatched regions indicate states outside the marginal ANN training input ranges.}}
    \label{fig:VLE_validation}
\end{figure}

\textcolor{\mycolor}{To further assess the thermodynamic accuracy of the ANN-aided VLE model, the predicted pressure and speed of sound are compared against reference VLE calculations, as shown in Figs.~\ref{fig:VLE_pressure} and \ref{fig:VLE_sound}. The comparison is performed for a nitrogen-dodecane mixture with mass fractions of $Y_{\mathrm{C_{12}H_{26}}}=0.01$ and $Y_{\mathrm{N_2}}=0.99$, covering temperatures from 270 to 1750 K and pressures from 0.5 to 14.5 MPa. These comparisons provide a direct assessment of the underlying ANN predictions against the reference thermodynamic model over a broad range of operating conditions.}



\textcolor{\mycolor}{As shown in Figs.~\ref{fig:VLE_pressure} and \ref{fig:VLE_sound}, the ANN predictions exhibit similar trends to the reference VLE calculations for pressure and speed of sound across most of the investigated domain. However, localized deviations are observed, particularly in regions of lower temperature where the thermodynamic properties exhibit greater sensitivity. The hatched regions indicate states outside the marginal training input ranges. These results demonstrate the predictive accuracy and limitations of the underlying ANN for the investigated thermodynamic properties, while the additional effects of ISAT interpolation and dynamic load balancing are evaluated through the subsequent CFD comparisons.}

To perform a qualitative comparison between the introduced methods, we compare contours of temperature, density, mass fraction of \ce{C12H26}, and the two-phase indicator $\alpha = \phi ( 1 - \phi)$. 
The results from the ISAT-ANN-aided VLE model, evaluated both with and without Dynamic Load Balancing (DLB), were compared against the benchmark ANN-aided model introduced by Srinivasan et al. \cite{srinivasan2024artificial}.
Section \ref{ISAT_ANN_metrics} will comment on the performance metrics of the ISAT-ANN-aided VLE model compared to the ANN-aided VLE model. The results described here utilize a static base mesh (1536 x 512 cells, \textcolor{\mycolor}{$\Delta x = 1.95 \mu m$}, 128 points across initial droplet diameter). The effects of adaptive mesh refinement are described in section \ref{AMR_metrics}.

\begin{figure}[htpb]
    \centering
    \includegraphics[width=\linewidth]{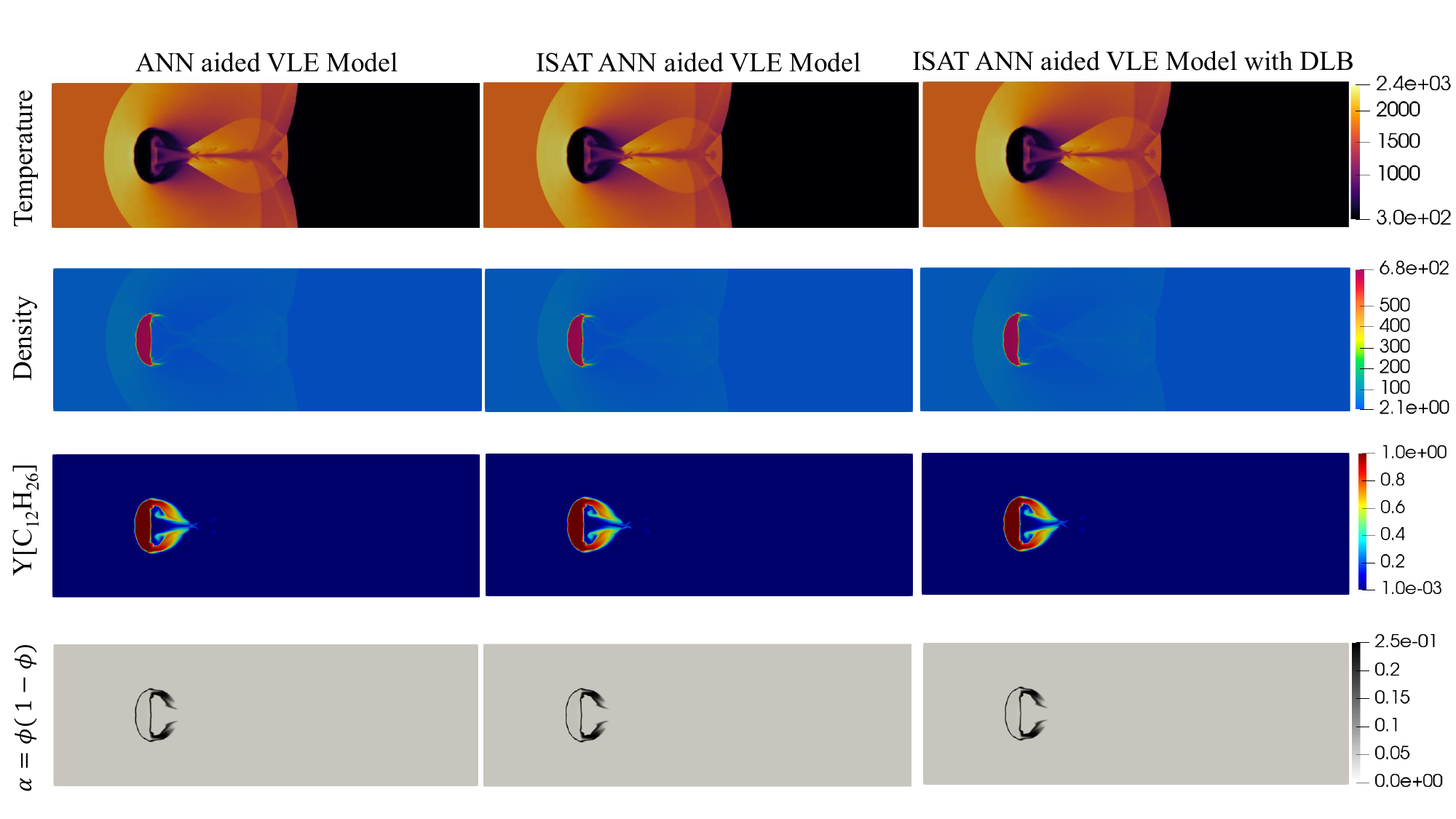}
    \caption{Comparison of temperature, density, mass fraction of \ce{C12H26}, and two-phase indicator $\alpha = \phi ( 1 - \phi)$  contours of the shock-droplet interaction case at time 0.08 $\mu s$ using ANN, ISAT-ANN, and DLB-aided ISAT-ANN models for VLE-based CFD.}
    \label{fig:contours}
\end{figure}

Figure \ref{fig:contours} illustrates the shock-droplet interaction at $t = 0.08$ $\mu s$. At this stage, the incident shock has traversed the droplet, inducing deformation and establishing both a stagnation bow shock and a trailing wake region. The ANN-aided VLE model serves as the benchmark to validate the ISAT-ANN and ISAT-ANN-DLB frameworks; the underlying validity of the ANN-aided VLE approach was established in a prior study \cite{srinivasan2024artificial}.

Visualization of the temperature field reveals a distinct bow shock characterized by a high-temperature stagnation zone. Subsequent interactions between the shock and the wake region generate secondary shocks, which eventually converge to form transmitted shock waves. Both ISAT-ANN and ISAT-ANN-DLB accurately capture these complex wave structures, exhibiting negligible deviations from the benchmark ANN-aided VLE model. Furthermore, density contours demonstrate consistent droplet deformation across all three methods, while mass fraction contours of \ce{C12H26} highlight high-vorticity regions within the wake. A significant advantage of this VLE-based CFD approach is the automatic identification of the interface. The two-phase indicator, $\alpha = \phi (1 - \phi)$, confirms that both ISAT-based models precisely locate the interface relative to the benchmark. Ultimately, this qualitative assessment confirms that neither the interpolative nature of ISAT nor the cross-rank local ISAT table communication inherent to DLB introduces discernible errors into the solution.

\begin{figure}[htbp]
    \centering
    \begin{subfigure}[b]{0.48\linewidth}
        \centering
        \includegraphics[width=\linewidth]{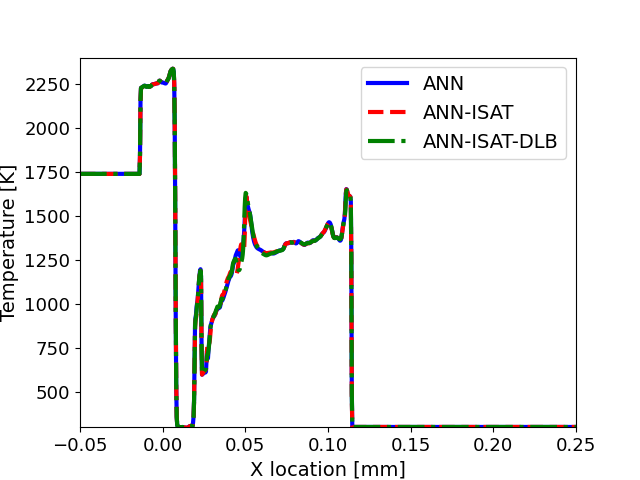}
        \caption{Temperature}
        \label{fig:T}
    \end{subfigure}
    \hfill
    \begin{subfigure}[b]{0.48\linewidth}
        \centering
        \includegraphics[width=\linewidth]{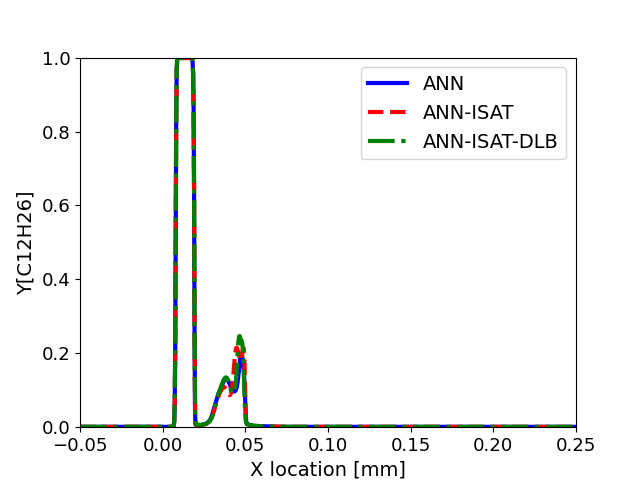}
        \caption{Y[\ce{C12H26}]}
        \label{fig:c12h26}
    \end{subfigure}
    \hfill
    \begin{subfigure}[b]{0.48\linewidth}
        \centering
        \includegraphics[width=\linewidth]{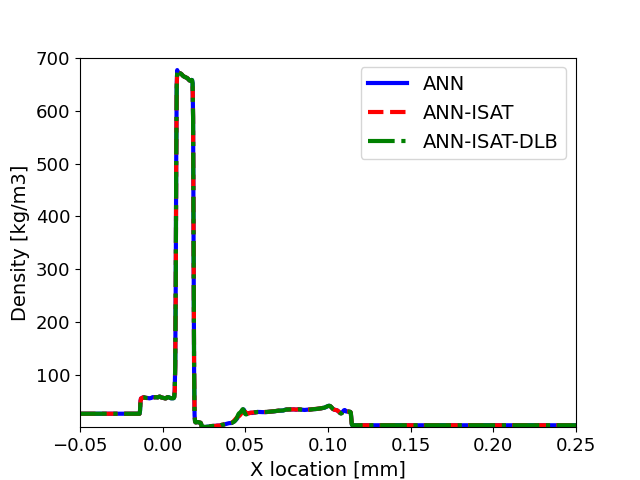}
        \caption{Density}
        \label{fig:rho}
    \end{subfigure}
    \hfill
    \begin{subfigure}[b]{0.48\linewidth}
        \centering
        \includegraphics[width=\linewidth]{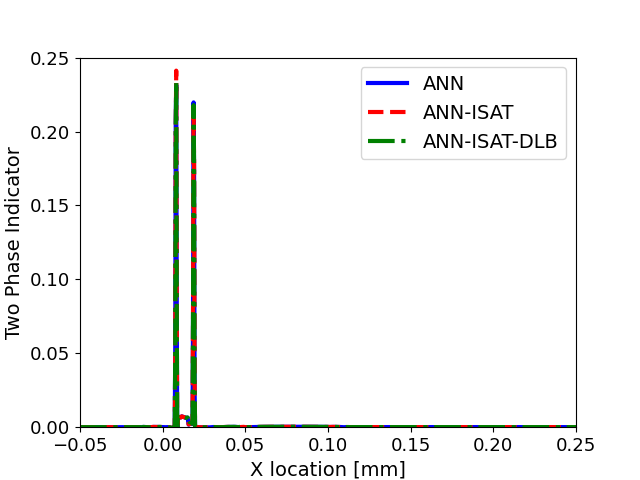}
        \caption{Two phase indicator $\alpha$}
        \label{fig:alpha}
    \end{subfigure}
    \caption{Comparison of centerline values of temperature, density, mass fraction of \ce{C12H26}, and two-phase indicator $\alpha = \phi ( 1 - \phi)$ of the shock droplet interaction case at time 0.08 $\mu s$ using ANN, ISAT-ANN, and DLB-aided ISAT-ANN models for VLE-based CFD.}
    \label{fig:quantitative}
\end{figure}

To quantitatively assess modeling errors, centerline profiles of temperature, density, \ce{C12H26} mass fraction, and the two-phase indicator, $\alpha = \phi(1 - \phi)$, are compared across the ANN, ISAT-ANN, and ISAT-ANN-DLB models at $t = 0.08$ $\mu s$ (Fig. \ref{fig:quantitative}). As shown in Fig. \ref{fig:T} and Fig. \ref{fig:rho}, the centerline values for temperature and density exhibit no discernible differences among the three methods. However, the \ce{C12H26} mass fraction profiles (Fig. \ref{fig:c12h26}) reveal minor deviations near $x = 0.05$ mm, corresponding to the wake region. These discrepancies are likely attributable to the high turbulence in this zone, which tends to amplify the inherent interpolative errors associated with ISAT-based tabulation. Similarly, the two-phase indicator (Fig. \ref{fig:alpha}) shows a marginal difference at the first peak located near the high-temperature stagnation zone at the droplet front. This slight variance likely stems from the sensitivity of vapor fraction predictions to fluctuations in internal energy, density, and mass fraction-the primary inputs for both the ANN and ISAT frameworks. Nevertheless, these differences remain minimal and do not lead to significant discrepancies in the overall results (as seen in fig. \ref{fig:contours}). Consequently, the observed error margins are deemed acceptable, particularly when weighed against the substantial computational speedup and enhanced performance metrics (detailed in section \ref{ISAT_ANN_metrics}) facilitated by the integration of ISAT and DLB.

\subsection{Computational Efficiency} \label{ISAT_ANN_metrics}
The primary motivation for integrating ISAT and DLB into the existing ANN-aided VLE model is to mitigate the high computational costs associated with simulating complex phenomena, such as a Mach 5 shock interacting with a dodecane droplet in the context of rotating detonation engines. To evaluate the computational efficiency of these techniques, particularly in parallel environments using multiple MPI ranks, the analysis is partitioned into two stages. First, the impact of ISAT on ANN-aided VLE modeling is assessed (Section \ref{isat_on_ann}), followed by an investigation into the effects of dynamic load balancing on the ISAT-ANN VLE framework (Section \ref{dlb_on_isatann}).

\subsubsection{Impact of ISAT on ANN aided VLE modeling} \label{isat_on_ann}
To evaluate the impact of ISAT on ANN-aided VLE modeling, a serial test case of the shock-droplet interaction (as described in Fig. \ref{domain}) was first conducted using a coarse mesh of $384 \times 128$ cells. This configuration corresponds to a spatial resolution of \textcolor{\mycolor}{$\Delta x = 7.81$ $\mu m$}, providing 32 points across the initial droplet diameter. The computational influence of ISAT was assessed using two primary metrics: CPU time and VLE time. CPU time represents the total duration required to complete a full time step, encompassing the solution of the governing equations, the VLE closure, and the estimation of all derived variables. In contrast, VLE time specifically isolates the duration required to solve the VLE equations (ANN inference or ISAT retrieval). \textcolor{\mycolor}{For parallel simulations, both metrics are evaluated using the maximum recorded time across MPI ranks at each timestep, with total execution times obtained by accumulating these timestep values over the simulation.} While CPU time captures the total simulation overhead, VLE time permits a more granular assessment of the efficiency gains achieved by the various modeling techniques introduced and therefore is the more rigorous metric for this comparative study.

\begin{figure}[htbp]
    \centering
    \begin{subfigure}[b]{0.48\linewidth}
        \centering
        \includegraphics[width=\linewidth]{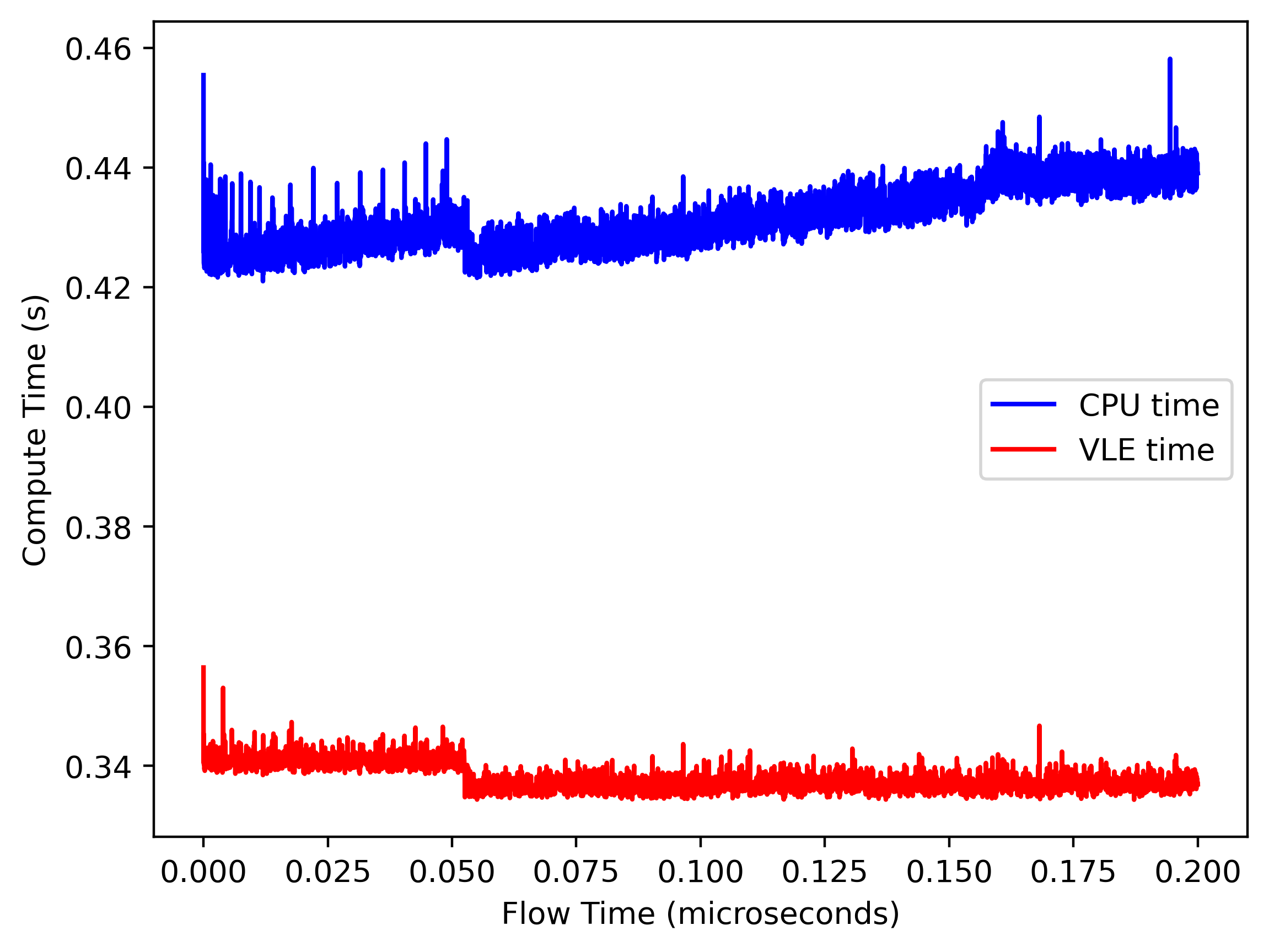}
        \caption{ANN-aided VLE model}
        \label{fig:ANNserial}
    \end{subfigure}
    \hfill
    \begin{subfigure}[b]{0.48\linewidth}
        \centering
        \includegraphics[width=\linewidth]{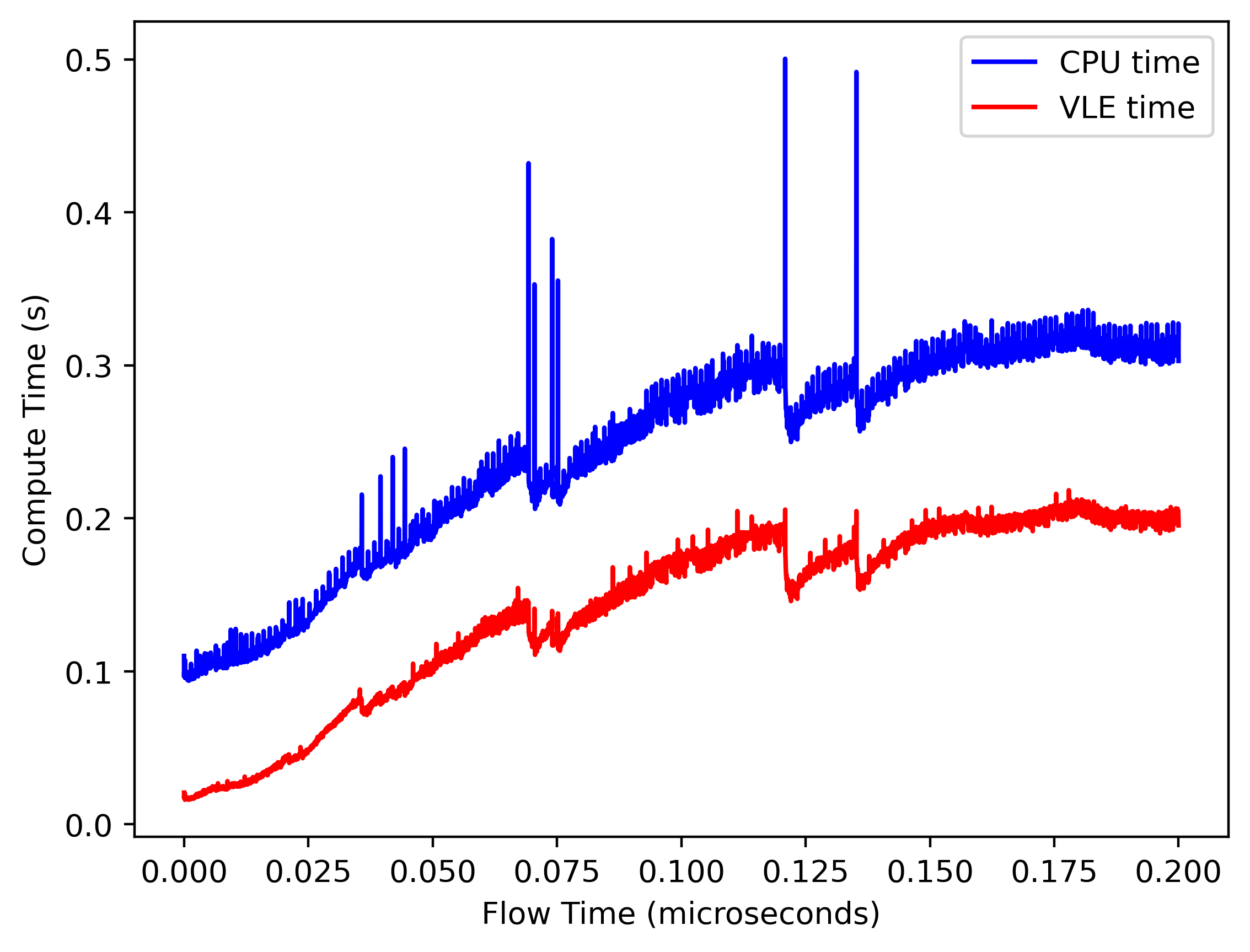}
        \caption{ISAT-ANN aided VLE model}
        \label{fig:ISATANNserial}
    \end{subfigure}
    \caption{Comparison of VLE time and CPU time during serial simulation of Mach 5 shock-droplet interaction between the ISAT-ANN VLE model and the ANN-aided VLE model using a coarse domain (384x128 cells).}
    \label{fig:serial}
\end{figure}

Figure \ref{fig:serial} presents a comparison of the CPU and VLE time consumed per time step for the ANN (Fig. \ref{fig:ANNserial}) and ISAT-ANN (Fig. \ref{fig:ISATANNserial}) models. A primary distinction is that the ANN-aided model maintains a nearly constant computational cost per step, whereas the ISAT-ANN model exhibits a progressive increase in execution time as the simulation advances. For the standalone ANN model, the VLE time remains stable at approximately 0.34 s, with a total CPU time of 0.43 s per timestep. Conversely, for the ISAT-ANN model, the VLE time scales from an initial 0.03 s to 0.2 s by the simulation's end, with a corresponding CPU time increase from 0.1 s to 0.3 s per timestep.

This temporal growth in the ISAT-ANN model is attributable to the expanding table size and the emergence of new thermodynamic states as the simulation advances. A larger table inherently increases retrieval time, while the occurrence of new states necessitates the insertion of new ISAT leaves, requiring both an ANN inference and a Jacobian evaluation. Despite this non-constant timing (a factor which introduces load imbalances), the ISAT-ANN model consistently outperforms the pure ANN approach. These results validate the primary objective of this study: integrating traditional tabulation techniques with modern neural networks to achieve significant computational performance gains.

\begin{table}[htpb]
    \centering
    \caption{Impact of In-Situ Adaptive Tabulation (ISAT) on ANN-aided VLE modeling using a single-core (serial) simulation of the Mach 5 shock-droplet interaction case using a coarse domain ($384 \times 128$ cells).}
    \begin{tabular}{|c|c|c|c|}
         \hline
         Model &  Total CPU time (s) & Total VLE time (s) & VLE Speed up\\
         \hline
         ANN & 2988.74 & 2338.34 & -\\
        \hline
         ISAT-ANN & 1721.32 & 1015.20 & 2.30\\
        \hline
    \end{tabular}
    \label{tab:singleCore}
\end{table}

The overall simulation runtime improvements are summarized in Table \ref{tab:singleCore}, which compares the total CPU and VLE execution times for the ANN and ISAT-ANN models. Despite the additional overhead associated with ISAT table management, the ISAT-ANN framework achieves speedup factors of $1.73\times$ for the total CPU runtime and $2.30\times$ for the VLE component. These gains are particularly impactful given that VLE calculations constitute approximately 79\% of the total computational cost in the baseline ANN-aided model and the introduction of ISAT-ANN successfully reduces this to 58\%. These promising results justify the overhead introduced by the ISAT algorithm. The subsequent performance of the ISAT-ANN model in multicore environments is further evaluated in section \ref{dlb_on_isatann}.

\subsubsection{Multicore Performance of ISAT-ANN - Impact of Dynamic Load Balancing} \label{dlb_on_isatann}

While the serial (single-core) performance of ISAT-ANN demonstrates clear advantages, these benefits do not necessarily translate directly to multicore environments, particularly in simulations utilizing the Message Passing Interface (MPI). In MPI-based parallelization, the computational domain is decomposed into smaller subdomains, each assigned to a specific CPU core (MPI rank). Each rank independently solves its assigned subdomain and communicates with adjacent ranks to exchange boundary cell information. This distributed memory architecture ensures that each rank operates within its own private memory space without direct access to the memory of other ranks. Translating this understanding of MPI-based simulations to ISAT-ANN-aided VLE modeling, we note that every rank maintains an identical instance of the ANN, but the associated ISAT tables typically diverge. This divergence occurs because each rank encounters a unique trajectory of thermodynamic states dictated by the local physics of its assigned spatial subdomain.

\begin{figure}[htbp]
    \centering
    \begin{subfigure}[b]{0.48\linewidth}
        \centering
        \includegraphics[width=\linewidth]{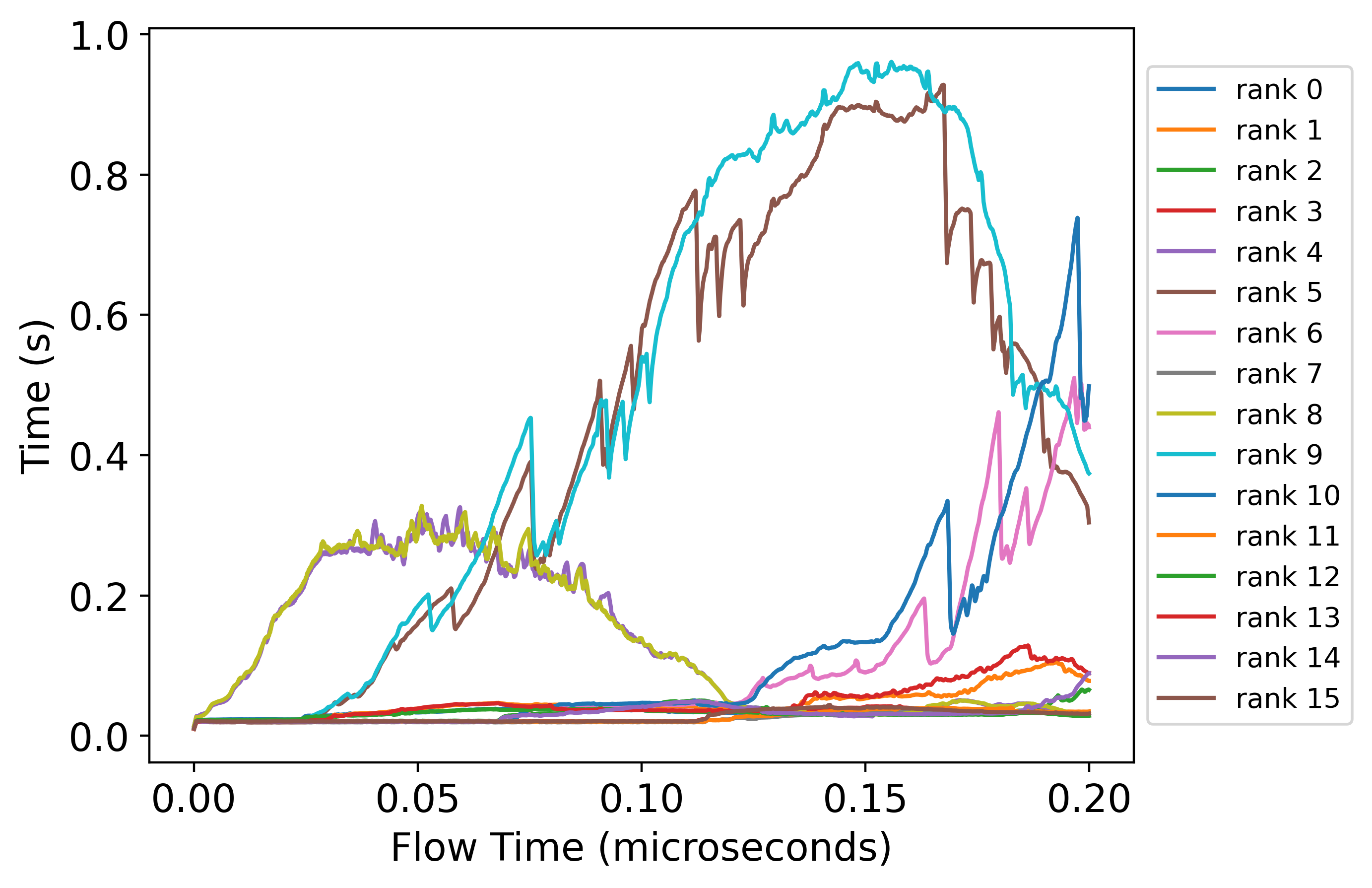}
        \caption{Without Dynamic Load Balancing}
        \label{fig:noDLB}
    \end{subfigure}
    \hfill
    \begin{subfigure}[b]{0.48\linewidth}
        \centering
        \includegraphics[width=\linewidth]{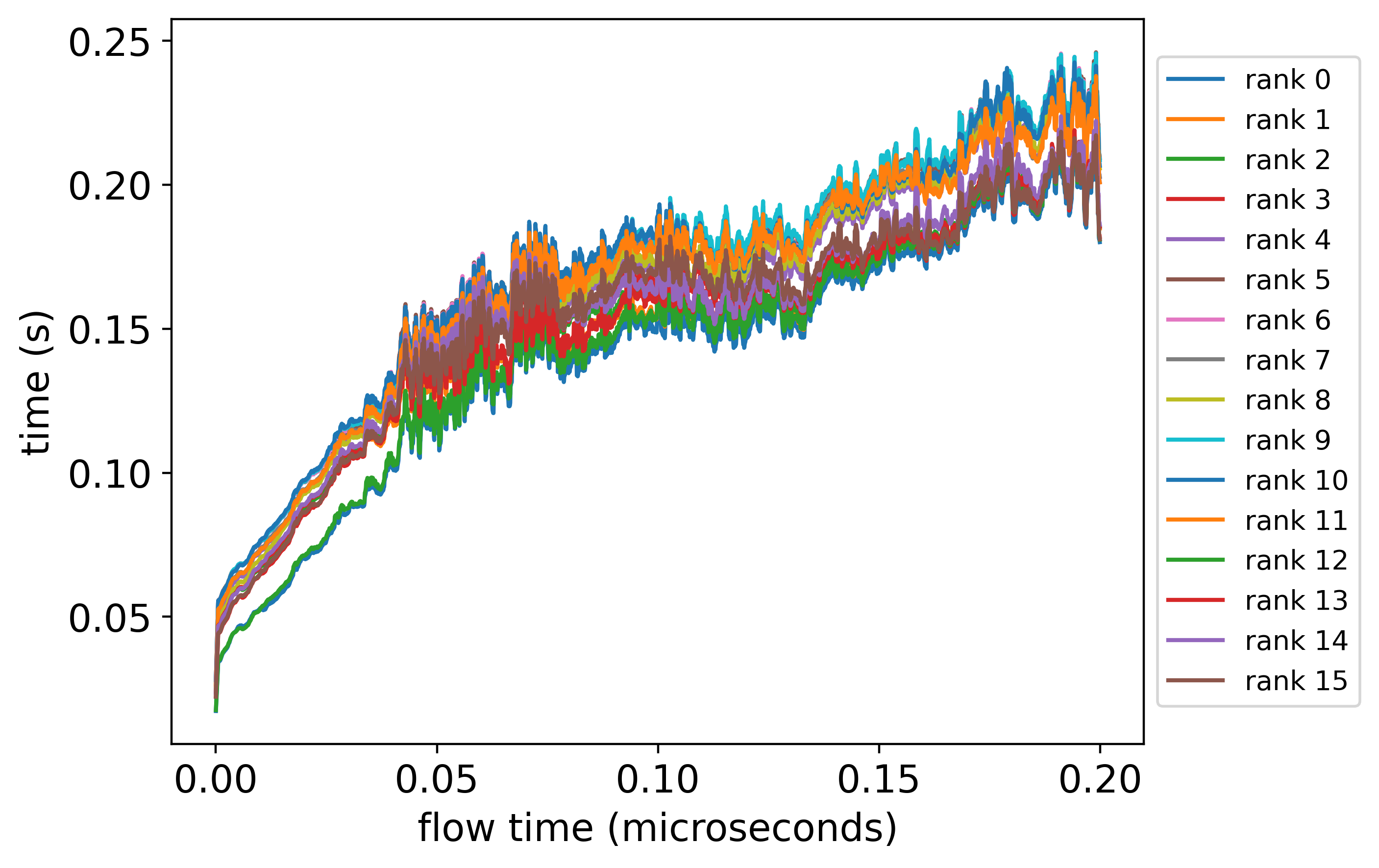}
        \caption{With Dynamic Load Balancing enabled}
        \label{fig:DLB}
    \end{subfigure}
    \caption{Comparison of VLE time during simulation time of ISAT-ANN VLE model with and without Dynamic Load Balancing (DLB)}
    \label{fig:DLBcomp}
\end{figure}

To evaluate the performance of ISAT-ANN in MPI-based parallel simulations, Figure \ref{fig:noDLB} illustrates the VLE execution time for a 16-core configuration. The computational domain is partitioned using an $8 \times 2$ decomposition, where the $x$-direction (normal to the shock) is divided into eight subdomains and the $y$-direction (parallel to the shock) into two. As shown in Figure \ref{fig:noDLB}, only two MPI ranks are active at any given time, while the remaining 14 ranks remain largely idle. This behavior is a direct consequence of the domain decomposition; as the shock wave traverses the domain, it occupies only two subdomains simultaneously, triggering the population of new thermodynamic states and high computational costs in those specific ranks. Conversely, ranks managing subdomains unaffected by the shock encounter minimal activity and complete their calculations rapidly. As the ISAT tables are localized within the private memory of each rank, thermodynamic data cannot be shared across the parallel environment without the introduction of additional computational techniques. Furthermore, memory addresses or references to an ISAT table are rank-specific and cannot be transferred via standard MPI communication. This localized architecture leads to severe load imbalances, where simulation progress is bottlenecked by a small subset of active ranks, resulting in inefficient resource utilization and significant performance degradation.

Dynamic Load Balancing (DLB) directly addresses the bottlenecks caused by localized computational loads by equitably distributing VLE calculations across all available MPI ranks. This distribution is determined by the computational time required to solve the VLE system for each individual CFD cell (detailed in Section \ref{DLB_VLE}). As illustrated in Figure \ref{fig:DLB}, the implementation of DLB ensures that all MPI ranks remain active throughout the simulation, facilitating optimal utilization of computational resources and maximizing throughput. Quantitatively, the maximum time per timestep for the ISAT-ANN model without DLB reaches approximately 1 s; however, integrating the DLB algorithm reduces this maximum to only 0.25 s. This significant reduction underscores the necessity and efficacy of DLB when employing ISAT-ANN within parallel computing environments.

Table \ref{tab:metric} presents a comparative analysis of the total CPU time, total VLE time, VLE computational percentage, and corresponding speed-up factors for the benchmark ANN, ISAT-ANN, and ISAT-ANN-DLB models utilizing 16 MPI ranks. A direct comparison between the benchmark ANN and the ISAT-ANN model reveals an increase in both CPU and VLE execution times for the ISAT-ANN model, resulting in a suboptimal speed-up factor of 0.91. This indicates that the increased overhead associated with rank-specific ISAT table construction and management effectively offsets the performance benefits of error-controlled tabulation in a parallel environment.

Conversely, the integration of Dynamic Load Balancing (DLB) with the ISAT-ANN framework achieves a speed-up factor of 2.62 relative to the benchmark ANN model, thereby facilitating the viable use of ISAT-ANN for MPI-based parallel simulations. The ISAT-ANN-DLB model further reduces VLE-related costs to 65.5\% of the total computational effort, a substantial decrease from the 82.1\% observed in the baseline ANN model. Notably, DLB is not anticipated to influence the performance of the standalone ANN-aided VLE model; since the ANN approach maintains a constant VLE execution time per step, it remains inherently load-balanced. This characteristic of the ANN-aided VLE framework was established in prior research \cite{srinivasan2024artificial}.

\begin{table}[htpb]
\centering
\caption{CPU time and VLE time comparison between ANN, ISAT-ANN and ISAT-ANN-DLB models for a 16 MPI ranks case of the Mach 5 shock-droplet interaction for a simulation time of 2e-7 s.}
    \begin{tabular}{|c|c|c|c|c|}
    \hline
         Model & CPU time (s) & VLE time (s) & VLE \% & VLE Speed up\\
         \hline
         ANN & 16184.53 & 13285.71 & 82.1\% & -\\
        \hline
        ISAT-ANN & 18690.11 & 14595.38 & 78.1\% & 0.91\\
        \hline
        ISAT-ANN-DLB & 7726.63 & 5063.28 & 65.5\% & 2.62\\
        \hline
    \end{tabular}
    \label{tab:metric}
\end{table}

The Dynamic Load Balancing (DLB) algorithm incorporates several sub-steps that contribute to the overall computational overhead of the VLE closure. Table \ref{tab:DLBsteps} details the execution time for each sub-step within the DLB framework (detailed in Section \ref{DLB_VLE}) for the shock-droplet interaction simulation utilizing 16 MPI ranks. Excluding the primary ``Solve'' phase, all stages represent administrative overhead necessary to ensure an equitable distribution of computational load. 

The ``Get Problems'' step, which involves packaging the thermodynamic state inputs into a dedicated ``VLE problem'' structure for each CFD cell, accounts for approximately 6\% of the total cost. This structure encapsulates essential state variables ($T, P, \phi, c$), cell IDs, and timing metrics. The ``Update State'' step, responsible for determining optimal workload distribution based on prior-step VLE timings and establishing the MPI communication schedule, comprises 15\% of the overall cost. Subsequently, the ``Balance'' step utilizes non-blocking MPI primitives to redistribute these problems according to the communication list, accounting for 8\% of the costs. The core ``Solve'' phase, wherein each rank processes both local and externally received VLE problems, constitutes 56\% of the total computational expenditure. Finally, the ``Unbalance \& Update'' stage employs non-blocking communication to return results to their respective parent ranks and updates the global CFD fields, such as temperature and pressure, within the OpenFOAM framework. 

Although the DLB-enabled ISAT-ANN framework demonstrates a substantial speed-up over the standard ANN-aided model, the cumulative overhead accounts for 44\% of the total computational cost. This significant overhead fraction may present scaling challenges as the number of MPI ranks increases. A comprehensive analysis of these scaling implications is provided in Section \ref{parallel_performance}.

\begin{table}[htpb]
    \centering
    \caption{Distribution of execution time among the different steps of the Dynamic Load Balancing algorithm for the Mach 5 shock--droplet interaction using 16 MPI ranks over a simulation time of $2\times10^{-7}$ s. Times represent the accumulated maximum across MPI ranks for each step.}
    \begin{tabular}{|c|c|c|}
    \hline
         DLB Algorithm Step & Time (s) & Percentage\\
         \hline
         Get Problems & 309.14 & 5.82\%\\
         \hline
         Update State & 792.15 & 14.92\%\\
         \hline
         Balance & 419.64 & 7.9\%\\
         \hline
         Solve & 2968.73 & 55.91\%\\
         \hline
         Unbalance \& Update &  819.39 & 15.43\%\\
    \hline
    \end{tabular}
    \label{tab:DLBsteps}
\end{table}

\textcolor{\mycolor}{Since the maximum execution time is accumulated independently for each DLB step across MPI ranks, the sum of these times can differ from the total VLE time in Table~2.} An analysis of the overhead associated with the DLB algorithm motivated further refinements to the underlying load-balancing logic, building upon the work of Tekgul et al. \cite {tekgul2021dlbfoam}. Specifically, the load distribution calculation within the "Update State" stage was modified to reduce computational complexity. By utilizing double-ended queues and a greedy allocation strategy-while simultaneously eliminating the sorting phase and redundant loops-the overall runtime of the Mach 5 shock-droplet simulation was significantly improved. Details regarding the modified DLB algorithm can be found in section \ref{DLB_VLE}. 

\begin{figure}[htbp]
    \centering
    \begin{subfigure}[b]{0.48\linewidth}
        \centering
        \includegraphics[width=\linewidth]{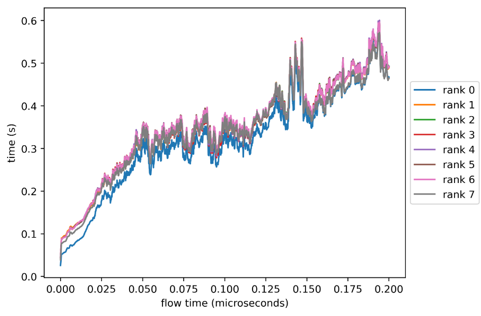}
        \caption{Original DLB algorithm}
        \label{fig:DLBorig}
    \end{subfigure}
    \hfill
    \begin{subfigure}[b]{0.48\linewidth}
        \centering
        \includegraphics[width=\linewidth]{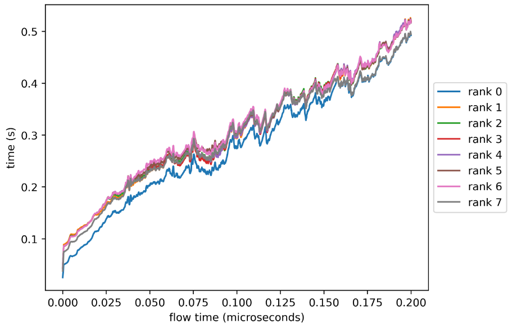}
        \caption{Modified DLB algorithm}
        \label{fig:DLBnew}
    \end{subfigure}
    \caption{Comparison of VLE time during simulation of Mach 5 shock-droplet interaction using the ISAT-ANN VLE model between the original DLB algorithm and the modified DLB algorithm. 8 MPI ranks have been utilized during this simulation.}
    \label{fig:DLBalgo}
\end{figure}

Figure \ref{fig:DLBalgo} compares the VLE time of the original and modified DLB algorithms utilizing 8 CPU cores for the Mach 5 shock droplet interaction case. The modified version exhibits more stable temporal behavior, with a peak VLE time of approximately 0.5 s per step, compared to the 0.6 s observed with the original algorithm. As summarized in Table \ref{tab:DLB_Modified}, these algorithmic refinements yield an additional 10\% speed-up in total CPU time and a 13\% improvement in VLE time.

\begin{table}[htpb]
    \centering
    \caption{CPU time and VLE time comparison between original and modified DLB algorithms for an 8 MPI ranks case of the Mach 5 shock-droplet interaction for a simulation time of $2\times10^{-7}$ s. Speedup factors (both CPU and VLE) are defined with respect to the original DLB algorithm.}
    \begin{tabular}{|c|c|c|c|c|}
    \hline
         DLB Algorithm &  CPU time (s) & VLE time (s) & CPU Speedup & VLE Speedup\\
         \hline
         Original & 14213.32 & 10722.77 & - & - \\
         \hline
         Modified & 12920.97 & 9488.34 & 1.10 & 1.13 \\
         \hline
    \end{tabular}
    \label{tab:DLB_Modified}
\end{table}

\subsection{Impact of Adaptive Mesh Refinement} \label{AMR_metrics}
As Adaptive Mesh Refinement (AMR) locally introduces new CFD cells, it inherently generates load imbalances due to the disproportionate distribution of cells across MPI ranks. Furthermore, as this spatial distribution can fluctuate rapidly during the simulation, performing a full domain re-decomposition each time AMR is triggered is computationally inefficient. The proposed DLB algorithm-which encapsulates thermodynamic states into a "VLE problem" structure and determines an optimal distribution strategy-effectively mitigates the imbalances introduced by AMR. This approach is particularly advantageous as it eliminates the necessity for repetitive and costly decomposition steps.

\begin{figure}[htbp]
    \centering
    \begin{subfigure}[b]{0.435\linewidth}
        \centering
        \includegraphics[width=\linewidth]{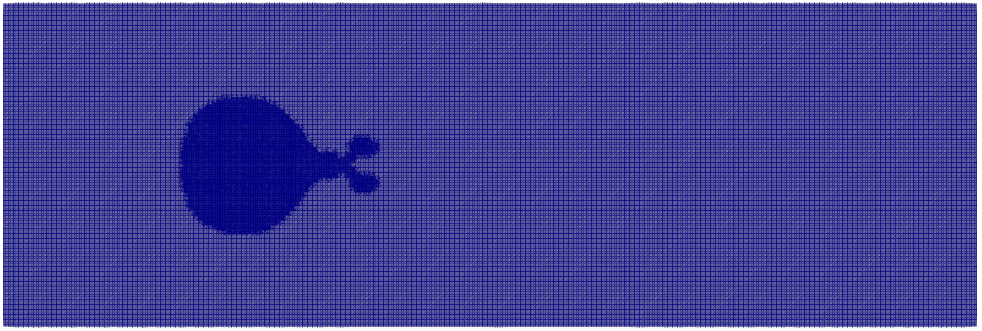}
        \caption{AMR modified CFD grid}
        \label{fig:amrsurface}
    \end{subfigure}
    \hfill
    \begin{subfigure}[b]{0.48\linewidth}
        \centering
        \includegraphics[width=\linewidth]{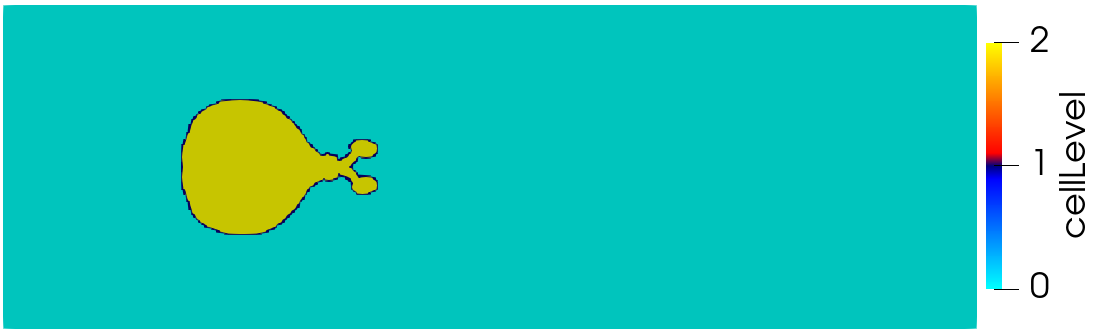}
        \caption{AMR Cell levels}
        \label{fig:cellLevel}
    \end{subfigure}
    \caption{AMR-enabled mesh visualized to showcase refinement occurring around the droplet. Cell levels (0: base, 1,2: refined) show the level of refinement. Level 1 is used as a transition level between level 0 (base) and level 2 (finest).}
    \label{fig:AMRsurfaceMesh}
\end{figure}

The ISAT-ANN-DLB algorithm was evaluated in conjunction with Adaptive Mesh Refinement (AMR) for the Mach 5 shock-droplet interaction. The static mesh configuration (without AMR) utilized the $1536 \times 512$ grid defined in Fig. \ref{domain}. In contrast, the AMR-enabled case employed a base mesh of $384 \times 128$, four times coarser than the static mesh, supplemented by two refinement levels. This configuration ensures that regions of interest maintain a spatial resolution equivalent to the static mesh. \textcolor{\mycolor}{The cell widths are $7.8125~\mu\mathrm{m}$, $3.90625~\mu\mathrm{m}$, and $1.953125~\mu\mathrm{m}$ at Levels 0, 1, and 2, respectively.} The mass fraction of C12H26 served as the refinement criterion, with thresholds established between \textcolor{\mycolor}{0.001 and 0.999}. Cells within this range were refined to the maximum level, while those outside were coarsened to the base grid. Additionally, the AMR algorithm implemented buffer cells to provide a smooth transition between disparate grid scales. As illustrated in Fig. \ref{fig:AMRsurfaceMesh}, the choice of trigger variable ensures that only the regions surrounding the droplet are resolved at the highest resolution. The resulting grid and the corresponding refinement levels are visualized in Fig. \ref{fig:amrsurface} and Fig. \ref{fig:cellLevel}, respectively. 
The base mesh (Level 0) and the finest mesh (Level 2) are bridged by Level 1 buffer cells; while the number of buffer layers is user-definable, \textcolor{\mycolor}{this study employs 1-4 buffer layers.}

\textcolor{\mycolor}{Across the 20 saved time directories, the global cell count ranged from 71,196 to 190,482, with a mean of 123,459.6 cells. At peak refinement, the mesh contained 39,360 Level 0 cells, 1,850 Level 1 cells, and 149,272 Level 2 cells, corresponding to an average of approximately 11,905 cells per MPI rank.}

\begin{figure}[htpb]
    \centering
    \includegraphics[width=\linewidth]{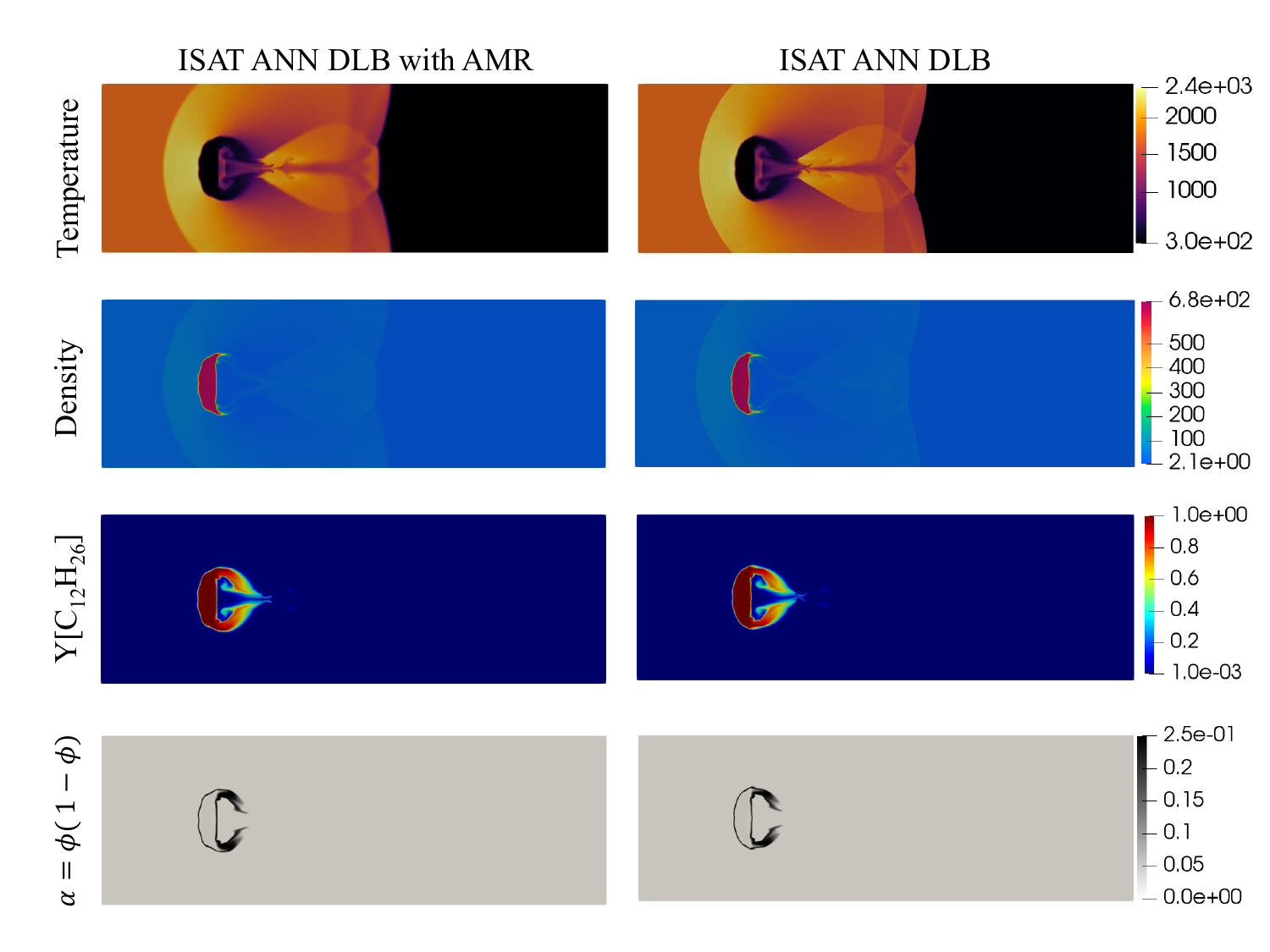}
    \caption{Comparison of temperature, density, mass fraction of \ce{C12H26} and two phase indicator $\alpha = \phi ( 1 - \phi)$  contours of the shock droplet interaction case at time 0.08 $\mu s$ between static mesh and adaptive mesh refinement (AMR) using ISAT-ANN-DLB VLE model.}
    \label{fig:ContoursAMR}
\end{figure}

Figure \ref{fig:ContoursAMR} presents a comparative analysis of the temperature, density, \ce{C12H26} mass fraction, and two-phase indicator ($\alpha = \phi(1 - \phi)$) for the Mach 5 shock-droplet interaction, comparing the ISAT-ANN-DLB model using a static mesh versus adaptive mesh refinement (AMR). As established, the AMR algorithm selectively refines the regions surrounding the droplet, dictated by the \ce{C12H26} mass fraction trigger, while leaving the shock wave regions at the coarser base resolution. Temperature and density contours reveal consistent droplet morphologies between the two methods. However, a closer examination of the temperature field indicates that while the global characteristics of the shock wave are preserved, the static mesh provides superior resolution of the shock front. Specifically, the static mesh yields a sharper shock wave profile due to the uniformly fine discretization compared to the unrefined AMR regions. Regarding the \ce{C12H26} mass fraction, both approaches capture similar droplet deformation and vorticity generation within the wake region. Finally, the two-phase indicator, $\alpha$, demonstrates that the predicted location of the liquid-vapor interface remains highly consistent across both meshing strategies.

\begin{figure}[htbp]
    \centering
    \begin{subfigure}[b]{0.48\linewidth}
        \centering
        \includegraphics[width=\linewidth]{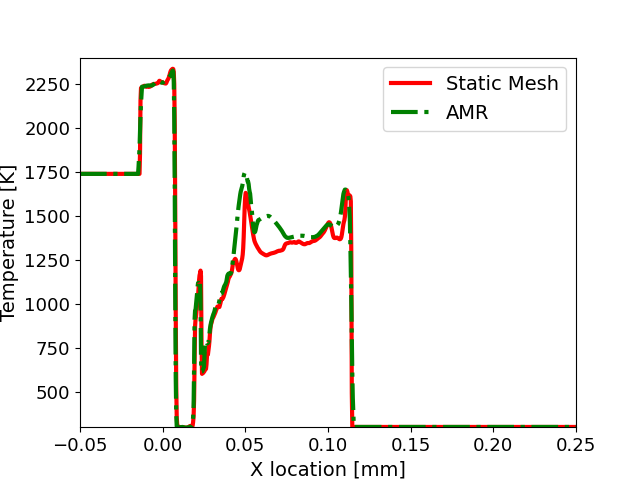}
        \caption{Temperature}
        \label{fig:T_AMR}
    \end{subfigure}
    \hfill
    \begin{subfigure}[b]{0.48\linewidth}
        \centering
        \includegraphics[width=\linewidth]{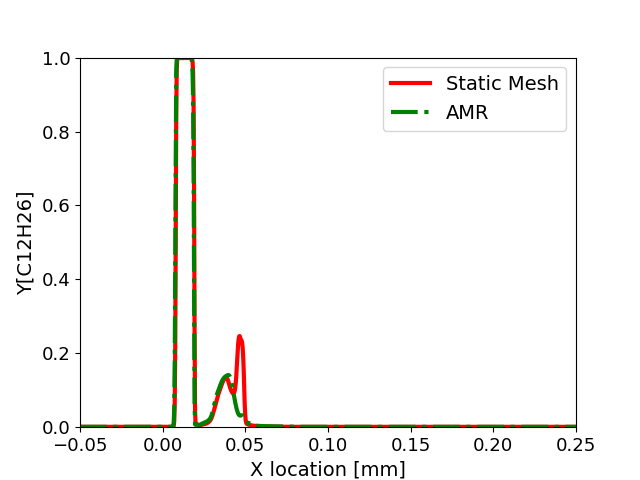}
        \caption{Y[\ce{C12H26}]}
        \label{fig:c12h26_AMR}
    \end{subfigure}
    \hfill
    \begin{subfigure}[b]{0.48\linewidth}
        \centering
        \includegraphics[width=\linewidth]{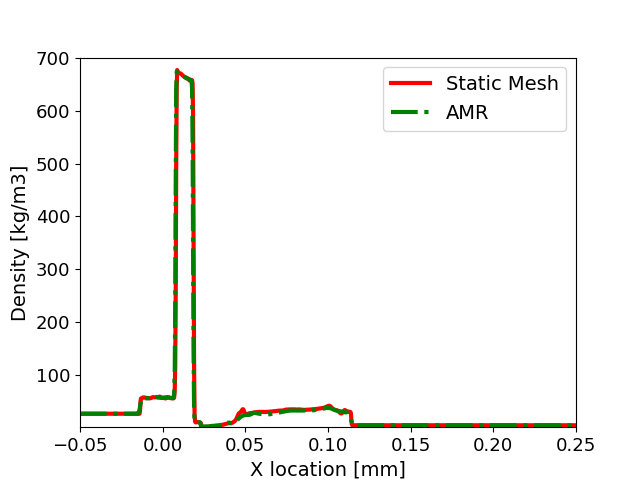}
        \caption{Density}
        \label{fig:rho_AMR}
    \end{subfigure}
    \hfill
    \begin{subfigure}[b]{0.48\linewidth}
        \centering
        \includegraphics[width=\linewidth]{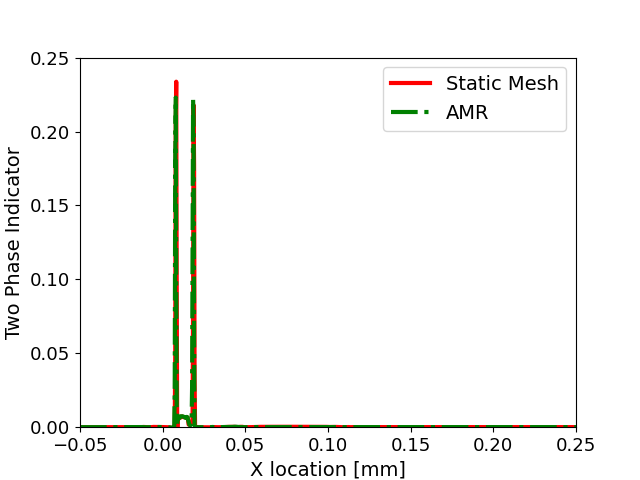}
        \caption{Two phase indicator $\alpha$}
        \label{fig:alpha_AMR}
    \end{subfigure}
    \caption{Comparison of centerline values of temperature, density, mass fraction of \ce{C12H26} and two phase indicator $\alpha = \phi ( 1 - \phi)$ of the shock droplet interaction case at time 0.08 $\mu s$ between static mesh and adaptive mesh refinement (AMR) using ISAT-ANN-DLB VLE model.}
    \label{fig:quantitative_AMR}
\end{figure}

To further evaluate the accuracy of the model, centerline values of temperature, density, mass fraction of \ce{C12H26}, and the two-phase indicator $\alpha = \phi ( 1 - \phi)$ are compared between the static and adaptive mesh methods in Fig. \ref{fig:quantitative_AMR}, similar to the analysis done in section \ref{qualitative}. Strong matches are seen between the static and AMR mesh cases in the regions before 0.01 mm (x location) in all the presented fields. Density and the two-phase indicator show a very good match throughout the entire plot. On the other hand, temperature and mass fraction of \ce{C12H26} start to show deviations close to x = 0.05 mm. These deviations can be attributed to the turbulent nature of the flow in the wake region, which can be impacted by mesh resolution. The overall matches observed between the static mesh and AMR-enabled mesh strengths demonstrate the applicability of AMR along with ISAT-ANN-DLB for the Mach 5 shock droplet case.

To quantitatively evaluate the accuracy of the adaptive mesh refinement approach, centerline profiles of temperature, density, \ce{C12H26} mass fraction, and the two-phase indicator, $\alpha = \phi (1 - \phi)$, are compared against the static mesh results in Fig. \ref{fig:quantitative_AMR}. This analysis follows the methodology established in Section \ref{qualitative}. 

Good agreement is observed between the static and AMR cases for all scalar fields in the upstream region ($x < 0.01$ mm). Furthermore, the density and two-phase indicator profiles exhibit a good match throughout the entire domain. In contrast, the temperature and \ce{C12H26} mass fraction profiles begin to show differences near $x = 0.05$ mm. These discrepancies are likely attributable to the inherently turbulent characteristics of the wake region, which are particularly sensitive to local mesh resolution. Despite these localized variations, the strong overall agreement between the static and AMR-enabled simulations underscores the viability of utilizing AMR in conjunction with the ISAT-ANN-DLB framework for Mach 5 shock-droplet interactions.

\begin{table}[htpb]
\centering
\caption{CPU time and VLE time comparison between static mesh and adaptive mesh refinement both using the ISAT-ANN-DLB model and 16 MPI ranks for the Mach 5 shock-droplet interaction case for a simulation time of 2e-7 s.}
    \begin{tabular}{|c|c|c|c|c|}
    \hline
         Model & CPU time (s) & VLE time (s) & VLE\% & VLE Speed up\\ 
        \hline
        Static Mesh & 7726.63 & 5063.28 & 65.5\% & -\\ 
        \hline
        AMR & 3537.73 & 1734.38 & 49.0\% & 2.92 \\ 
        \hline
    \end{tabular}
    \label{tab:AMRmetric}
\end{table}

While the accuracy of utilizing AMR has been established through comparison with static mesh results, the primary motivation for its introduction and implementation was to further mitigate the computational burden of VLE-based CFD simulations. Table 5 compares the total CPU and VLE execution times for a 16-rank MPI simulation of the Mach 5 shock-droplet case, comparing the static and AMR-enabled approaches under the ISAT-ANN-DLB framework. The AMR-enabled case achieves \textcolor{\mycolor}{a speed-up factor of 2.18$\times$ in total CPU time and 2.92$\times$ in VLE computation time} over the static mesh counterpart, justifying the integration of AMR into VLE-based CFD. In the static mesh configuration, VLE calculations account for 65.5\% of the total computational cost; however, this proportion is reduced to 49\% when AMR is employed. It is important to note that this reduction in the VLE cost percentage may not solely be attributed to a decrease in the number of VLE calculations, but likely also reflects the additional overhead associated with AMR operations, such as local grid refinement and coarsening.

\subsection{Parallel Performance} \label{parallel_performance}

The integration of the ISAT-ANN framework alongside computational strategies such as Dynamic Load Balancing (DLB) and Adaptive Mesh Refinement (AMR) is intended to mitigate the significant computational costs associated with VLE-based CFD, particularly in parallel environments. While parallelization can be implemented through shared-memory (e.g., OpenMP) or distributed-memory (e.g., MPI) paradigms, the present analysis focuses exclusively on the latter. This choice is dictated by the architecture of the OpenFOAM framework, which serves as the development platform for the discussed techniques.

\begin{figure}[htpb]
    \centering
    \includegraphics[width=0.6\linewidth]{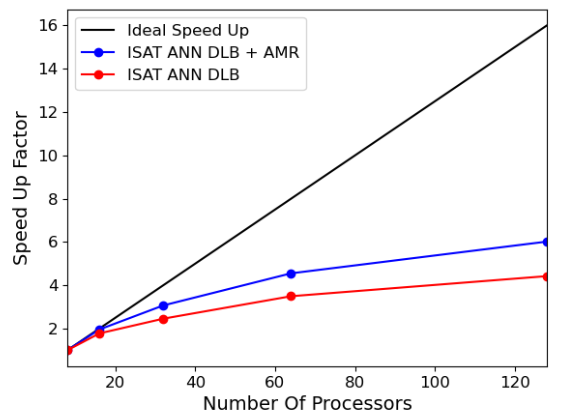}
    \caption{Strong parallel scaling of ISAT-ANN-DLB with and without AMR.}
    \label{fig:scaling}
\end{figure}

A strong scaling study was conducted for the Mach 5 shock-droplet interaction (domain illustrated in Fig. \ref{domain}) to evaluate the performance acceleration of the ISAT-ANN-DLB model as computational resources are increased. Both a static mesh ($1536 \times 512$ cells) and an AMR-enabled configuration ($384 \times 128$ base mesh with two refinement levels) were evaluated independently, with results presented in Fig. \ref{fig:scaling} comparing the observed speed-up against the theoretical ideal. While both the static and AMR-enabled ISAT-ANN-DLB models exhibit near-ideal scaling between 8 and 16 processors, the performance begins to diverge from the linear trend beyond 16 processors. This degradation in parallel efficiency is attributed to two primary factors: communication overhead and a decrease in the number of cells per MPI rank. 

\begin{figure}[htpb]
    \centering
    \includegraphics[width=\textwidth]{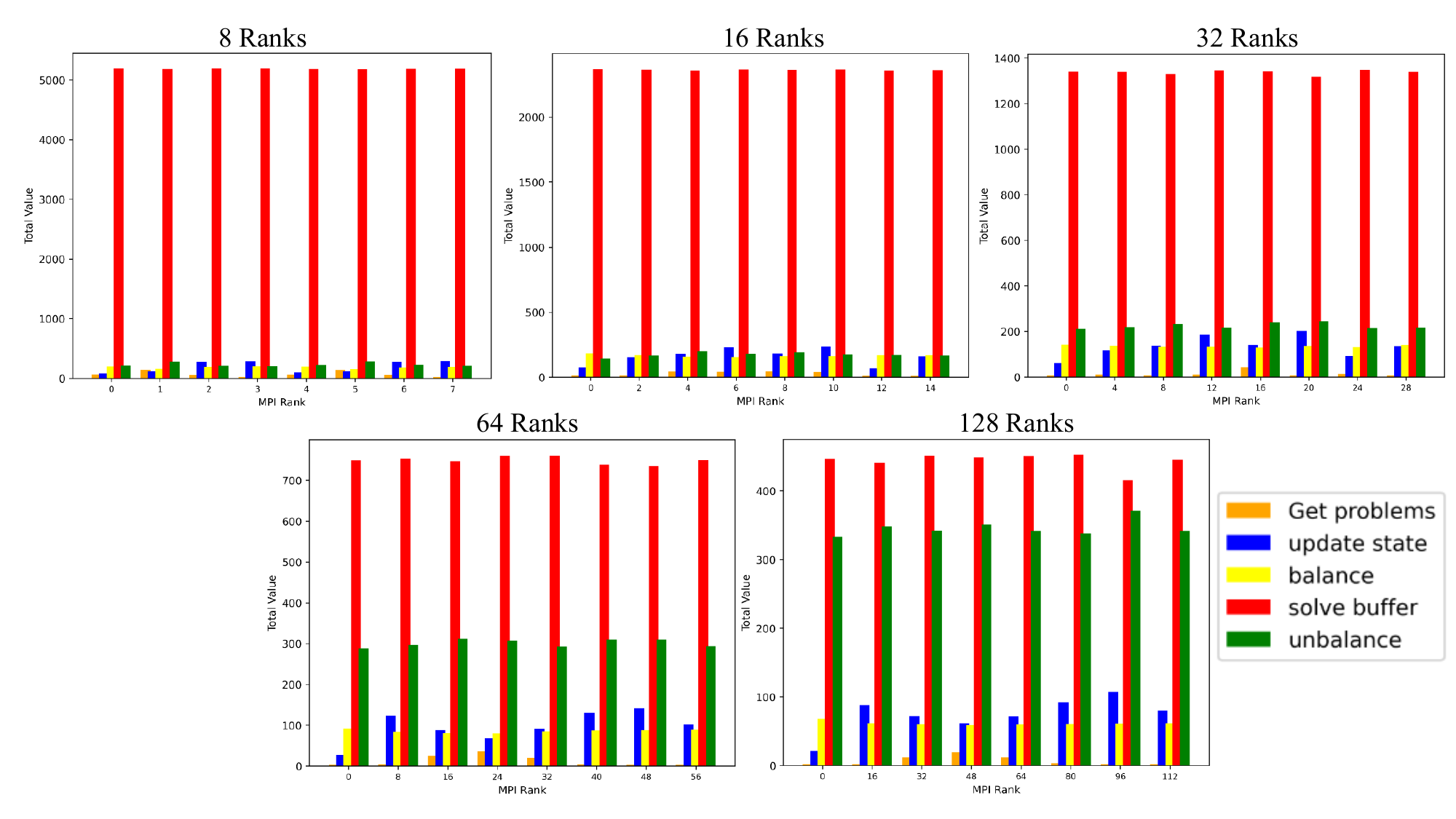}
    \caption{Distribution of computational time among the various operations of the dynamic load balancing step. For ideal scaling to be achieved, maximum amount of time must be spent in the solve buffer step (red).}
    \label{fig:parallel_costs}
\end{figure}

First, the global communication and administrative overhead inherent in DLB operations increases significantly with the number of MPI ranks. To characterize this increase, Fig. \ref{fig:parallel_costs} illustrates the distribution of computational costs for each DLB stage across varying rank counts. For the DLB algorithm to maximize computational gains and justify its inherent overhead, the "solve" phase (indicated by the red bar in the figure) must remain the dominant component of the runtime. As observed, the administrative overhead becomes significant at 32 ranks and beyond, leading to the observed scaling bottlenecks.

Second, as the number of processors increases for a fixed global mesh, \textcolor{\mycolor}{the average number of CFD cells owned per rank decreases. This can reduce the computational savings relative to the overhead of table management, ANN inference, and communication. However, the actual thermodynamic workload also depends on ISAT hit rates, workload distribution, and the redistribution of VLE queries through DLB.} In this study, the static mesh contains a total of 786,432 cells; at 16 ranks, this results in approximately 49,152 cells per rank. \textcolor{\mycolor}{The near-ideal scaling observed up to 16 ranks, followed by reduced efficiency at 32 ranks, identifies a case-specific operating point rather than a general minimum cell-count requirement.} Interestingly, the AMR-enabled case demonstrates slightly superior scaling characteristics compared to the static mesh counterpart. This is likely due to the localized, rank-specific nature of AMR operations, which execute concurrently across the domain. The presence of this rank-local AMR overhead reduces the relative dominance of the more communication-intensive DLB and VLE calculations in the overall runtime profile. Finally, while domain decomposition strategies (e.g., $8 \times 2$ vs. $4 \times 4$ partitioning) are known to influence MPI communication costs and scaling performance, an exhaustive analysis of decomposition topology was beyond the scope of this work.

\section{Conclusions} \label{conclusions}

High-pressure transcritical and supercritical flows play a pivotal role in modern propulsion and power-generation technologies, including liquid-propellant combustors, scramjets, and rotating detonation engines. Accurately modeling these systems requires capturing complex real-fluid thermodynamics and phase change through multicomponent VLE calculations. However, conventional iterative thermodynamic solvers impose a major computational bottleneck, often consuming more than 90\% of total simulation runtime and becoming prone to convergence failure under steep thermodynamic gradients.

This work addresses these limitations through an integrated ISAT-ANN framework that combines the complementary strengths of neural-network inference and adaptive tabulation. The ANN provides \textcolor{\mycolor}{non-iterative} state predictions during table expansion, while ISAT avoids repeated network evaluation through rapid local retrieval. To extend these gains to large-scale parallel simulations, an O(N) greedy dynamic load-balancing strategy redistributes the strongly nonuniform thermodynamic workload across MPI ranks. Adaptive mesh refinement is further integrated to concentrate spatial resolution \textcolor{\mycolor}{around the droplet interface and regions of deformation} without uniformly refining the computational domain. The resulting framework was evaluated using a demanding \textcolor{\mycolor}{nominal} Mach 5 shock-droplet interaction involving a $C_{12}H_{26}$ dodecane droplet in an $N_2$ background at \textcolor{\mycolor}{a prescribed initial post-shock pressure of} 145 bar.

Qualitative and quantitative evaluations demonstrate that the ISAT-ANN framework, with and without dynamic load balancing, preserves the predictive accuracy of the previously validated benchmark ANN approach \cite{srinivasan2024artificial} while accurately capturing complex shock structures and droplet evolution. The ISAT-ANN approach also provides substantial acceleration in serial simulations. In larger parallel simulations, however, unbalanced local tabulations degraded performance to a $0.91\times$ speedup because of severe workload imbalance across MPI ranks. The greedy load-balancing algorithm resolved this bottleneck and restored the computational advantage of the hybrid approach. When combined with adaptive mesh refinement, the fully coupled solver continued to significantly outperform the standalone ANN solver. \textcolor{\mycolor}{Scaling analysis showed near-ideal efficiency up to 16 MPI ranks for the investigated configuration, corresponding to approximately 49,152 cells per rank, with diminishing efficiency at higher rank counts due to increasing computational and communication overheads.}

Taken together, these results establish the proposed method as a promising computational platform for extreme multiphase flow environments involving complex thermodynamics, strong spatial gradients, and highly nonuniform workloads. As a newly developed architecture, the present implementation represents an initial realization of this approach, with further improvements in tabulation strategy, load redistribution, network integration, and parallel communication expected to provide additional gains in robustness and efficiency. More broadly, the underlying architecture may be extended to other computationally intensive multiphysics applications, including detailed chemical kinetics, plasma dynamics, and radiation transport. Future work will focus on further optimizing the framework and extending it to fully three-dimensional turbulent reacting flows with detailed chemical kinetics and more complex multiphase configurations.

\section*{Acknowledgements}
S. Yang gratefully acknowledges the support from the Office of Naval Research (ONR) grant under Award No. N00014-22-1-2287 under the supervision of project monitors Dr. Steven Martens and Dr. David Gonzalez. The authors gratefully acknowledge the computing resources provided by the DoD High Performance Computing Modernization Program (HPCMP) and Minnesota Supercomputing Institute (MSI).

\section*{AUTHOR DECLARATIONS}
\subsection*{Conflict of Interest Statement}
The authors have no conflicts to disclose.
\subsection*{Authorship contribution}
\textbf{Navneeth Srinivasan:} Conceptualization, Data curation, Formal analysis, Investigation, Methodology, Software, Validation, Visualization, Writing - original draft. 
\textbf{Po-Yu Lai:} Methodology, Writing - review \& editing.
\textbf{Yousef Eldahshoury:} Methodology, Writing - review \& editing.
\textbf{Suo Yang:} Conceptualization, Formal analysis, Funding acquisition, Investigation, Methodology, Project administration, Resources, Supervision, Writing - review \& editing.

\section*{Data availability statement}
The data that support the findings of this study are available from the corresponding author upon reasonable request. 

\appendix
\bibliographystyle{elsarticle-num}
\bibliography{ref}
\end{document}